\pdfoutput=1 
\documentclass[pre,twocolumn,showpacs,amsmath,amssymb,floatfix,eqsecnum]{revtex4}

\usepackage{graphicx}
\usepackage{times}

\begin{document}

\newcommand\CLH[1]{{\bf CLH:#1}}
\newcommand{\MEMO}[1]{{\sl #1}}
\newcommand{\SAVE}[1]{}    
\newcommand{\OMIT}[1]{}
\newcommand{\LATER}[1]{}
\newcommand{\MFL}[1]{{\bf MFL: #1}}

\def \beq {\begin{equation}}
\def \eeq {\end{equation}}
\def \beqa {\begin{eqnarray}}
\def \eeqa {\end{eqnarray}}

\newcommand{\QQ} {{\bf Q}}

\newcommand{\tab}{\hspace*{2em}}
\newcommand{\mb}{\mathbf}
\newcommand{\ul}{\underline}
\newcommand{\ahat}{\hat{\mathbf{A}}_0}
\newcommand{\bhat}{\hat{\mathbf{B}}_0}
\newcommand{\chat}{\hat{\mathbf{C}}_0}
\newcommand{\xhat}{\widehat{\mathbf{e}}_1}
\newcommand{\yhat}{\widehat{\mathbf{e}}_2}
\newcommand{\zhat}{\widehat{\mathbf{e}}_3}
\newcommand{\Ham}{\mathcal{H}}
\newcommand{\Esite}{\mathcal{E}_i}
\newcommand{\Eavg}{\bar{\mathcal{E}}}
\newcommand{\JLT}{\tilde{J}}
\newcommand{\QLT}{\mb{Q}_{L.T.}}
\newcommand{\LTeig}{\lam_{L.T.}}
\newcommand{\al}{\alpha}
\newcommand{\be}{\beta}
\newcommand{\lam}{\lambda}
\newcommand{\un}{\underline}
\newcommand{\dg}{^{\circ}}
\newcommand{\pr}{\partial}
\newcommand{\what}{\widehat}

\title{Ground States of the Classical Antiferromagnet on the Pyrochlore Lattice}

\author{Matthew F. Lapa$^{1,2}$ and Christopher L. Henley$^{1}$}
\affiliation{$^{1}$Laboratory of Atomic and Solid State Physics, Cornell University,
Ithaca, New York, 14853-2501 \\
$^{2}$Department of Physics, University of Illinois at Urbana-Champaign, 61801-3080}

\date{\today}

\begin{abstract}

We study the classical ground states of the exchange-coupled Heisenberg antiferromagnet on the Pyrochlore lattice, a non-Bravais
lattice made of corner-sharing tetrahedra. In particular, we map out the entire phase diagram for the case of first and second
nearest neighbor interactions. In this phase diagram we find {\it four} complex non-coplanar ground states based on different ordering
modes. These are the Cuboctahedral Stack state, a $\langle111\rangle$ stacking of Cuboctahedral states, and three families of
commensurate spiral: the Kawamura states, constructed from three different combinations of $\{\tfrac{3}{4}\tfrac{3}{4}0\}$ 
modes, the Double-Twist state, also constructed from $\{\tfrac{3}{4}\tfrac{3}{4}0\}$ modes, and the Multiply-Modulated 
Commensurate Spiral state, constructed from $\{\tfrac{3}{4}\tfrac{1}{4}\tfrac{1}{2}\}$ modes. We also briefly look at states 
involving the two kinds of third nearest neighbor interactions on the Pyrochlore lattice.
In this region of parameter space we again find the Cuboctahedral Stack state, 
and we also find another non-coplanar state in the form of a new kind of Alternating Conic Spiral.

\end{abstract}


\maketitle

\section{Introduction}

\LATER{0 Used ``widehat'' in $\widehat{\mb{m}}$ but maybe use plain
hat in $\hat{\mb{n}}$.}

This paper concerns the classical ground states of the Heisenberg Hamiltonian
on the Pyrochlore lattice, a non-Bravais lattice made of corner-sharing tetrahedra.
In this paper we focus our analysis on the case of first and second nearest 
neighbor interactions $J_1$ and $J_2$, and we briefly mention some results 
that include the effects of third nearest neighbor interactions $J_3$ and $J_3'$.
(Each site on the Pyrochlore lattice has two distinct kinds of 
third nearest neighbor, at the same separation, yet inequivalent by symmetry --
having distinct neighbors at the same distances appears to be universal 
in non-Bravais lattices.)

An important motivation to work out the phase diagram of magnetic systems is
the inverse problem arising from neutron diffraction: not the crystallographic
problem of fitting the magnetic structure from the measured Bragg intensities,
but rather the physical problem of inferring -- in the absence of magnon
dispersion relations that would require dynamic neutron scattering -- 
for what possible interactions is the observed structure the ground state.
(One motivation for this project was the puzzle presented in Ref. \cite{Ma08}
in GeNi$_2$O$_4$, where it was necessary
to posit interactions as far as $J_4$.)
\SAVE{FOR A future paper which actually includes
$J_4$).
This was plausible, from the viewpoint of the Goodenough-Kanamori 
superexchange rules, for the exchange paths in this material 
and the oxidation states of the constituent cations~\cite{Ma08}.  
However, the ground state of the suggested couplings was not actually
determined in Ref.~\onlinecite{Ma08}, and thus  it is unsettled
whether the proposed couplings there actually explain the experimentally
discovered state.
This last state is stacked along a $\{111 \}$ axis, and thus
consists of two alternating layers with different densities of spins.
As we noted [elsewhere], such a stacked spin state
has a good chance to develop a non-coplanar spiral.
}
\SAVE{
The question of which couplings
might stabilize the observed order
arose for some $B$-spinels.
In the case of CdCr$_2$O$_4$~\cite{Chu05}, {\it no}
exchange Hamiltonian seems to suffice, which was a
clue that other physics is in play, namely
magnetoelastic interactions~\cite{Tch02,Tch04b,Che06}.
}

The other motivation is to facilitate the theorists' pursuit of model systems 
with unusual properties.
For example, whenever frustrated magnets have a large number of degenerate 
ground states, quantum or thermal fluctuations, or dilution, can select a 
particular ground state 
--  a form of so-called ``ordering due to disorder''~\cite{henley89}.  
Thus, one interest in our study is to identify parameter combinations
on the phase diagram that lead to such degeneracies.  
This tends to happen along
``degenerate phase boundaries'' (see Ref.~\onlinecite{SH}, 
sections II C and VII): their property is that one gets two
limiting states depending on which side of the boundary the limit is taken from,
yet it is possible to continuously turn one of these into the
other within the manifold of degenerate states that exists on the boundary.

Another special property, which we are particularly concerned with in
this paper, is inherently {\it non-coplanar} states.
These can produce an {\it anomalous Hall effect} due to
Berry phases of itinerant electrons \cite{Ta01,kalitsov08,taillef06,anomHall};
they open up possibilities of {\it multiferroic} 
behavior~\cite{cheong-mostovoy,khomski09,kaplan-CoCr2O4};
they turn into {\it chiral} spin liquids if spin order
is destroyed; and their symmetry-breaking leads to an 
order parameter in the form of an $O(3)$ matrix, which 
allows unusual topological defects ($Z_2$ vortex and 
$Z_2$ domain wall~\cite{henley84a}).

The Pyrochlore spin lattice represents many antiferromagnets in 
any of three well-known crystal structures: the B sites of the 
spinel structure, one cation sublattice (of the two equivalent ones)
in the Pyrochlore structure; or 
the sublattice of small atoms in the (metallic)
Laves phase structure.
It is highly degenerate when only the nearest-neighbor 
interaction is included; in studies where other terms were added that
break the degeneracy, surprisingly few of them added 
farther-neighbor coupling (apart from dipolar spin ice~\cite{spin-ice}).
Instead,  anisotropies were added such as 
local easy planes~\cite{bramwell-easy-plane} or
Dzyaloshinskii-Moriya anisotropic exchange~\cite{DM}.
Alternatively, lattice distortions
were considered~\cite{Tch02,Tch04b,Che06,bergman-lattice}.
\LATER{1 Add more citations on Er2Ti2O7 easy plane!}

In the case of the Pyrochlore, Ref.~\cite{Rei91} worked out
(in effect) the optimum Luttinger-Tisza modes for general Hamiltonians
with $J_1$ up to $J_4$, but limited to $J_3=J_3'$.
\LATER{2 Mention the other limitations...}
The main systematic exploration to date
has been of the $(J_1,J_2)$ Pyrochlore [or the $(J_1,J_3)$ case,
which is equivalent to the $(J_1,J_2)$ case so long as $J_1$
is large and antiferromagnetic in sign].
It uncovered  a multiple-wavevector, not quite commensurate state
with especially soft fluctuations \cite{Tsu07,Na07,Che08,Oku11}
that is not quite fully understood.
Another context in which farther neighbor couplings naturally
arise is in metallic Pyrochlores, in the form of electron-mediated
R.K.K.Y. exchange interactions (that oscillate and decay with 
separation as a power law).
\LATER{3 Add Kawamura spin glass and Motome.}

\LATER{4 CLH needs to make a pass through various papers I have
copies of, to make sure they are cited, as well as to look
through the so-far uncited references at back.  This can be
delayed till LATER.}

\section{Notations and Methods}

In this section we describe the Pyrochlore lattice and we also explain the methods and diagnostics we used to generate and analyze the spin configurations. We present a number of ways of visualizing the Pyrochlore lattice, as certain ways of thinking about the structure of the Pyrochlore lattice are helpful for understanding certain spin configurations. We also discuss the Iterative Minimization algorithm, our main method of generating spin configurations. Finally we discuss a number of ways of understanding and interpreting the results of the Iterative Minimization simulation, including the Luttinger-Tisza method, numerical Fourier transforms, least square fits and Variational optimization.

\subsection{Pyrochlore Lattice}

The Pyrochlore lattice consists of four face-centered cubic (F.C.C.) sublattices, numbered 0, 1, 2 and 3. We take the edge-length of the cubic unit cell of each F.C.C. sublattice to be one. We choose the origins of these four F.C.C. sublattices to be located at $(0,0,0)$, $(0,\tfrac{1}{4},\tfrac{1}{4})$, $(\tfrac{1}{4},0,\tfrac{1}{4})$ and $(\tfrac{1}{4},\tfrac{1}{4},0)$, respectively.

\begin{figure}[b]
  \caption{The Pyrochlore lattice can be viewed as a stacking of two-dimensional triangular and Kagome lattice layers along a $\langle111\rangle$ axis. The picture shows one (black) Kagome lattice layer sandwiched between an upper (red) triangular lattice layer and a lower (blue) triangular lattice layer.}
\vskip 10pt
  \centering
    \includegraphics[width=0.5\textwidth]{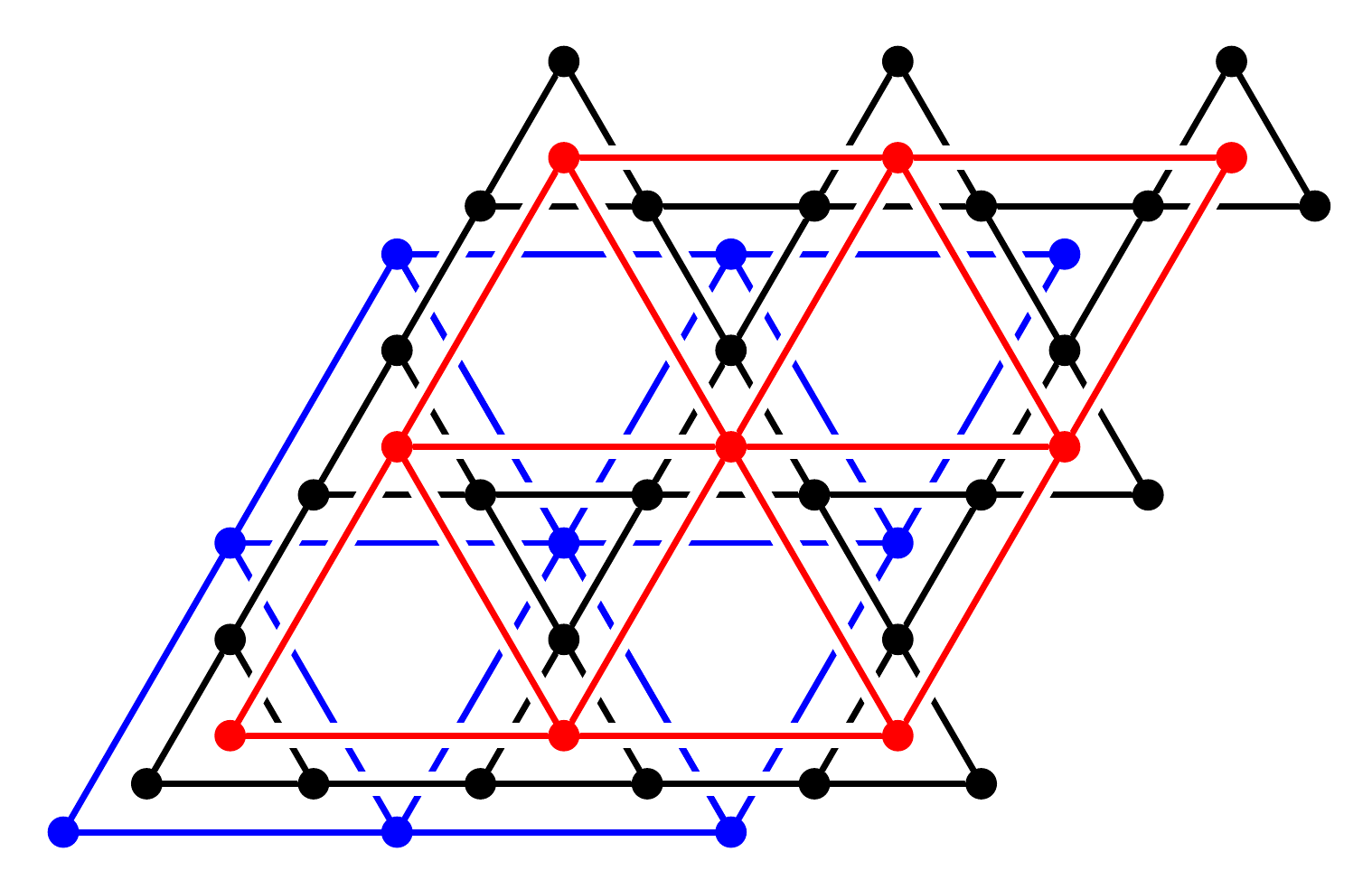}
\label{fig:KagTriangle}
\end{figure}

Alternately, the Pyrochlore lattice can be viewed as an ABC stacking of triangular and Kagome lattice layers along a $\langle111\rangle$ direction in real space as in figure \ref{fig:KagTriangle}.

Our Hamiltonian is
	\beq
		\Ham = -\frac{1}{2}\sum_{i=1}^{N}\sum_{j\neq i}J_{ij}\mb{s}_i\cdot\mb{s}_j\ , \label{HeisHam}
	\eeq
The $\mb{s}_i$ are classical 3-component unit vectors which sit on the sites $\mb{r}_i$ of the Pyrochlore lattice. The lattice contains $N$ sites. A positive (negative) $J_{ij}$ represents a ferromagnetic (antiferromagnetic) interaction between the spins $\mb{s}_i$ and $\mb{s}_j$. Since the interaction $J_{ij}$ is a function of the displacement vector $\mb{R}_{ij} = \mb{r}_i - \mb{r}_j$ between the spins $\mb{s}_i$ and $\mb{s}_j$, we can write $J_{ij} = J(\mb{R}_{ij})$. 

Each site on the Pyrochlore lattice has six first nearest neighbors, 12 second nearest neighbors and 12 third nearest neighbors, but the third nearest neighbors come in two kinds, and each site has six of each kind. First nearest neighbors have an interaction $J_1$, second nearest neighbors have an interaction $J_2$, third nearest neighbors of the first kind have an interaction $J_3$ and third nearest neighbors of the second kind have an interaction $J_3'$. From now on we refer to first nearest neighbors as ``$J_1$ neighbors", second nearest neighbors as ``$J_2$ neighbors", etc.

The $J_1$ neighbors of a site can be reached by traveling along one edge of a tetrahedron. The $J_2$ neighbors of a site can be reached by traveling along two non-collinear edges of two adjacent tetrahedra. The $J_3$ neighbors of a site can be reached by traveling along two collinear edges of two adjacent tetrahedra. Finally, the $J_3'$ neighbors of a site can be reached by hopping across a hexagon in the Kagome lattice layers. The two kinds of third nearest neighbors are obviously inequivalent by symmetry since every two $J_3$ neighbors have a spin lying halfway between them on the line connecting them, while there is no spin lying on the line connecting two $J_3'$ neighbors.

In this rest of this paper we limit our study to $J_1$ and $J_2$, 
except that $J_3$ and $J_3'$ are used in Section~\ref{CS} and Appendix~\ref{ACS}, 
and $J_3$ is mentioned in Section~\ref{qEquals0}.

\subsection{Luttinger-Tisza Analysis}

\LATER{5 ? does anything need to be inserted.
MFL wrote, I am still having some trouble explaining why and how the 
LT method goes wrong. I'm not sure how to justify the idea that the weights 
of the LT modes on the different sublattices are given by the corresponding 
eigenvector of the LT matrix. 
CLH: Was email sent 9/25/12 sufficient? 
MFL: I think it is good for now (10/24). }

\LATER{6  add a remark about this?
You and Sophia too much emphasized the list of wavevectors;
What matters equally is which mode at that wavevector, i.e. 
how the amplitudes and relative complex phases go with the 
different sublattices = sites in the unit cell.
I gave an example in discussing the cuboc states in Sophia's
paper.  Incidentally that was another good example of
diametrically opposite states that use the same wavevectors...
Could a short remark or footnote about this be inserted 
somewhere?  Does it belong in the LT section???}

The Luttinger-Tisza method~\cite{LT,Kapl06} is a method for finding the ordering wavevector that characterizes the lowest energy state of \eqref{HeisHam} for a given set of interactions $J_{ij}$. Recall that the goal of any method (analytical or computational) for finding the ground states of \eqref{HeisHam} is to find the set of $N$ spins $\{\mb{s}_i\}$ which minimize the total energy subject to the constraint that all of the spins have unit length, i.e. $\left|\mb{s}_i\right|^2 = 1$ $\forall$ $i$. The constraint that all of the spins have unit length is often called the \textit{strong constraint}.

The Luttinger-Tisza method attempts to accomplish this by instead trying to minimize \eqref{HeisHam} subject to what is known as the \textit{weak constraint},
	\beq
		\sum_{i}\left|\mb{s}_i\right|^2 = N\ . \label{WC}
	\eeq

By formulating the problem in terms of the Fourier Transforms of the spin configuration in each Bravais sublattice, one can obtain a lower bound on the energy of the true ground state. One then searches for a solution to the weak constraint problem which is composed solely of modes related by the symmetry of the underlying Bravais lattice. In an ideal situation we would be able to find a solution to the weak constraint problem that also satisfies the strong constraint, is composed solely of modes related by the symmetry of the underlying Bravais lattice, and has an energy equal to the lower bound. If such a configuration could be found, it would rigorously be the ground state.

\SAVE{CLH Notation: 
Use a(n abuse of) notation ``$i\in \alpha$'' meaning spin $i$ is in sublattice $\alpha$.
And a notation $\alpha(i)$ to mean ``the sublattice that spin $i$ belongs to.''
Here we write ``$J_{\alpha\beta}(\mb{R})$'',
where $\mb{R}$ is the vector between sites.}

To formulate the method we first rewrite equation \eqref{HeisHam} in a way which clearly shows which spins are in which Bravais sublattice. We use the notation $\al(i)$ to indicate which Bravais sublattice spin $i$ belongs to, so the possible values for $\al(i)$ are 0, 1, 2 and 3 in our case. When it is convenient we will also write $i \in \al$ to indicate that spin $\mb{s}_i$ is located in the $\al^{th}$ Bravais sublattice. From here on we specialize to the case of the Pyrochlore lattice, which has four F.C.C. sublattices. 
Then we can write
	\beq
		J_{ij} \equiv J_{\al(i)\be(j)}(\mb{R}_{ij})
        \label{JijLT}
	\eeq
where $J_{\al\be}(\mb{R})$ is the interaction between spins in  F.C.C. sublattices $\al$ and $\be$ that
are separated by a vector $\mb{R}$.

Next we define the Fourier Transform of the spin configuration in the $\al^{th}$ Bravais sublattice to be
	\beq
		\tilde{\mb{S}}_{\al}(\mb{q}) = \frac{1}{\sqrt{N/4}}\sum_{i\in\alpha} \mb{s}_i e^{-i\mb{q}\cdot\mb{r}_i}\ , 
	\eeq
where $\mb{q}$ is a wavevector in the Brillouin zone of the F.C.C. lattice and $N/4$ is the number of spins 
in each F.C.C. sublattice. The inverse transform is
	\beq
     \mb{s}_i  = \frac{1}{\sqrt{N/4}}\sum_{\mb{q}} 
e^{i\mb{q}\cdot\mb{r}_i} \tilde{\mb{S}}_{\al(i)}(\mb{q})\ .
	\eeq
Substituting this into \eqref{JijLT} 
\SAVE{(and making use of the usual orthogonality relations for discrete Fourier transforms)}
yields the result
	\beq
		\Ham = -\sum_{\mb{q}}\sum_{\al,\be} \JLT^{\al\be}(\mb{q})\tilde{\mb{S}}_{\al}(\mb{q})\cdot\tilde{\mb{S}}_{\be}(-\mb{q})\ . 
\label{HeisQ}
	\eeq

The coefficients of the quadratic form \eqref{HeisQ} form 
a $4\times4$ matrix $\ul{\JLT}(\mb{q})$ with matrix elements
	\beq
	\JLT_{\al\be}(\mb{q}) \equiv 
        \frac{1}{2}\sum_{j\neq i}J_{\al(i)\be(j)}(\mb{R}_{ij})e^{i\mb{q}\cdot\mb{R}_{ij}} .
        \label{LTelements}
	\eeq
Let $\lam_{\nu}(\mb{q})$ be the four eigenvalues of this matrix at wavevector $\mb{q}$, 
each with a normalized eigenvector ${u}^{\nu}_\alpha(\mb{q})$, 
where $\alpha$ runs over the four sublattices. 
These modes, transformed back to the actual lattice as $u^{\nu}_{i} \equiv u^{\nu}
(\mb{r}_i)$, are exactly analogous to 
the Bloch wavefunction for the $\nu$ band.
Express the spin configuration in terms of these eigenmodes:
        \beq
               \tilde{\mb{S}}_\al(\mb{q}) = \sum_\nu \tilde{\mb{w}}_\nu u^\nu_\al(\mb{q}),
        \eeq
so the coefficients have three Carteisan components for spin. 
The total energy is then 
	\beq
		\Ham = -\sum_{\mb{q}}\sum_\nu \lam_{\nu}(\mb{q}) \left|\tilde{\mb{w}}_{\nu}(\mb{q})\right|^2\ .
	\eeq

\SAVE{A confusing notation for $\tilde{\mb{w}}(\mb{q})$ is used for this vector in Sklan \& Henley.}

Let $\lam_{max}(\mb{q})$ be largest eigenvalue of the matrix $\ul{\JLT}(\mb{q})$ and let $\QLT$ be a wavevector 
that maximizes $\lam_{max}(\mb{q})$ (generically there is a star of symmetry-related wavevectors with magnitude $|\QLT|$ that all do this). 
The quantity $\LTeig \equiv \lam_{max}(\QLT)$ is called the optimal Luttinger-Tisza eigenvalue,  
\SAVE{Both notations $\LTeig$ and $\lam_{max}$ are handy. 
Later on in the paper there are a few times where MFL compares
the largest eigenvalues of the J matrix for a few suboptimal wavevectors to the largest eigenvalue for the LT wavevector, and I was thinking that using both $\lam_{max}(\mb{q})$ and $\LTeig$ would help me do that in a concise way).}
and $\QLT$ is the optimal Luttinger-Tisza wavevector. We refer to $\un{\JLT}(\QLT)$ as the Luttinger-Tisza matrix.

In Fourier space the weak constraint \eqref{WC} becomes
$\sum_{\mb{q},\nu}|\tilde{\mb{w}}_{\nu}(\mb{q})|^2 = N$ so the ground state energy manifestly satisfies the inequality
	\beq
		\Ham \geq -N\LTeig \label{Ebound} \ .
	\eeq 

A necessary condition for achieving this minimum is that the three components of $\tilde{\mb{S}}_\al(\mb{q})$ be linear combinations of the Luttinger-Tisza eigenmodes $u_{\al}^{\nu}(\QLT)$. The Luttinger-Tisza method works if one can find a normalized spin configuration composed solely of eigenmodes of $\QLT$ (and symmetry-related wavevectors) which also achieves the lower bound on the energy, $-N\LTeig$. On non-Bravais lattices, it is not always possible to construct a normalized state using only Luttinger-Tisza eigenmodes. Many states require admixtures of a set of suboptimal modes in order to achieve normalization \cite{SH}.  Our experience working with the Pyrochlore lattice confirms that this is true.

 \SAVE{There may be no way to add different (symmetry-related) modes to make a state which is normalized in every sublattice. 
In this case one needs to incorporate other suboptimal wavevectors to construct a normalized state. There may be multiple ways to admix other wavevectors to construct a normalized state which can lead to extra degeneracies in the ground state. }


\SAVE{There may be multiple ways to admix other wavevectors to construct a normalized state which can lead to extra degeneracies in the ground state. }

The key difficulty with the Luttinger-Tisza method is to ensure $|\mb{s}_i|\equiv 1$ for every site, when that
is not generally the case for a Luttinger-Tisza eigenmode (expressed in real space, where it appears as a plane wave with a certain weight on each sublattice). On Bravais lattices, that {\it is the case} if the eigenmodes are written as complex
plane waves $\exp(i\QLT\cdot \mb{r})$; from this we can always construct a ground state in the
form of a coplanar spiral by using one component for the sine and another for the cosine~\cite{Kapl06}. 
Since the optimal Luttinger-Tisza wavevectors typically occur in symmetry-related stars, a nontrivial degeneracy
would be possible by taking linear combinations of two modes with different $\mb{Q}$'s from the
same star;  however, since each mode requires two spin components, that is not possible
with physical three-component spins.  But whenever $\QLT$ is half of a reciprocal lattice
vector (and thus lies on a special point of the Brillouin zone boundary), we have $\exp(i\QLT\cdot\mb{r})=\pm 1$.
Thus the plane-wave modes are {\it real} in this case, and we can combine up to three of them,
associated with orthogonal spin directions, to construct continuous families of degenerate
classical ground states.

\SAVE{We cannot use different Q's for the four different sublattices, 
so that we end up with a degenerate family of ground states made from more than just three symmetry-related Q's.
The spin directions going with each wavevector need to be perpendicular, so you
would need four-component spins.}

\SAVE{It is only when we can make a ground state
exactly from LT modes that we can hope for nontrivial degeneracies.  
Otherwise, the admixing typically breaks the degeneracy.}

\LATER{7 MFL: Should I include the figure of the LT phase diagram for J1-J2 as well as the real phase diagram?
TODO: CLH think it should be included, but the right place for it is not 
immediately obvious to me.
If it's very similar to Figure~\ref{fig:PhaseDiag}, I considered maybe LT phase diagram
could be indicated by adding additional dotted lines to that figure.  But that's probably 
a bad idea.)}

For each set of couplings $J_{ij}$, we can calculate the optimal Luttinger-Tisza wavevector $\QLT$. 
This allows us to create a ``Luttinger-Tisza phase diagram", showing in what domains of parameter
space each kind of L.T. mode is optimal.
The L.T. phase diagram tells us the expected wavevector content of the ground state for a given set of couplings, 
but cannot tell how those modes are combined to construct a normalized state, 
whether there are multiple degenerate ways to do so,
or whether the actual ground state requires admixtures of suboptimal modes for normalization. 
So although the Luttinger-Tisza phase boundaries typically correspond closely to the real phase boundaries,
the true phase diagram often has extra subdivisions that the LT phase diagram lacks, e.g.
the three Kawamura states (which all use the same Luttinger-Tisza modes, with admixtures of suboptimal modes). 
\SAVE{The Luttinger-Tisza phase diagram merely tells us that in that region of parameter space 
$\QLT= 2\pi(\frac{3}{4},\frac{3}{4},0)$.}

\subsection{Iterative Minimization method}

Following Ref.~\onlinecite{SH}, our main method for finding the ground states of \eqref{HeisHam} 
is the Iterative Minimization simulation, a numerical approach which starts with a 
random spin configuration and generates states with progressively lower energy, until the method converges
to some (local) energy minimum.
Let us first define the local field $\mb{h}_i$ to be minus the gradient of \eqref{HeisHam} 
with respect to the components of spin $i$ (analog of the force in a
mechanical system):
	\beq
 		\mb{h}_i \equiv  -\frac {\delta \Ham}{\delta \mb{s_i}} = \sum_{j\neq i} J_{ij}\mb{s}_j \ .
	\eeq
In any ground state, every spin must be aligned parallel to its local field. 
(If any spin were not, reorienting it along its local field would immediately lower the energy.)
\SAVE{This property is the basis of the Iterative Minimization method.}

The method works in the following way. We start with $N$ spins on the Pyrochlore lattice, all pointing in random directions. Next we run a large loop, and on each iteration of the loop we pick $N$ spins at random (one at a time) and reorient them so that they point along their local field. 
\SAVE{Since we pick the $N$ spins at random during each iteration, it is possible that some spins are reoriented multiple times and other spins are not rotated at all during any given iteration of the loop. We stop the simulation when the spin configuration stops changing appreciably after each successive iteration (in practice, when each spin component changes by less than some small number $\epsilon$ for many successive iterations of the loop).}

In our simulations we mostly work on a lattice that is an $L_x \times L_y \times L_z$ block of Pyrochlore lattice unit cells (where $L_x$, $L_y$ and $L_z$ are integers) with periodic boundary conditions. We also have the capability to use twist boundary conditions, in which spins near a certain boundary of the lattice see their neighbors beyond that boundary twisted by an angle $\theta$ about a certain axis in spin space.

Once we have obtained a spin configuration using Iterative Minimization, we try to get a sense of the basic structure of the state using a few simple diagnostics. These are the Site Energy diagnostic, which gives a rough sense of the layout of the spin configuration in real space, the Common-Origin plot, which indicates which directions the spins are pointing in in spin space, and numerical Fourier Transforming, which gives the wavevector content of the state. Finally, we use Least Square fits of the spin configuration to the wavevectors appearing in the numerical Fourier Transform to get an approximate parameterization for the ground state. For simpler states, we can then use the technique of Variational Optimization to determine the idealized values of the parameters appearing in our approximate parameterization of the state.
(All of these except the site energy were introduced in Ref.~\onlinecite{SH}.)

\subsubsection{Site energy diagnostic}

We define the site energy of the spin $\mb{s}_i$ to be
	\beq
		\Esite= -\frac{1}{2}\sum_{j\neq i}J_{ij}\mb{s}_i\cdot\mb{s}_j\ . \label{siteE}
	\eeq
The average of this quantity over all the spins in the lattice is the ``energy per site" and we denote it by $\Eavg \equiv \tfrac{1}{N}\sum_i \Esite$. The total energy of the system can then be written as $\Ham = N\Eavg$. 

We often find that spin configurations break up into sublattices in real space according to the site energy of the spins in those sublattices. Sometimes these sublattices are the familiar four F.C.C. sublattices of the Pyrochlore lattice, as in the Cuboctahedral Stack state (see section \ref{CS}), but sometimes the spin configuration breaks up into a more exotic set of sublattices, as in the Kawamura Sextuplet-q state (see section \ref{KawamuraSq}).

\subsubsection{Fourier Transforming}

Once we have found a spin configuration using Iterative Minimization, we take Fourier Transforms of that spin configuration in each F.C.C. sublattice. Next we compute the Spin Structure Factor in each F.C.C. sublattice to determine which wavevectors the configuration mostly consists of. The Fourier Transform of the spin configuration in the $\al^{th}$ F.C.C. sublattice is
	\beq
		\tilde{\mb{S}}_{\al}(\mb{q}) = \frac{1}{\sqrt{N/4}}\sum_{i\in\alpha} \mb{s}_i e^{-i\mb{q}\cdot\mb{r}_i}  
	\eeq 
where $\mb{q}$ is a wavevector in the Brillouin zone of the F.C.C. lattice and the sum is taken over all spins $\mb{s}_i$ in the $\al^{th}$ F.C.C. sublattice. In practice we compute these Fourier Transforms one spin component at a time. 

We then examine $\left|\tilde{\mb{S}}_{\al}(\mb{q}) \right|^2$ for each F.C.C. sublattice (here the square includes a sum over
the three Cartesian components from the spins, as well as real/imaginary parts from the Fourier transform.)
Sorting these in order of decreasing weight identifies the main
wavevectors that characterize the state, which can then be checked against the list of
optimal Luttinger-Tisza wavevectors to help determine whether the spin configuration found from
Iterative Minimization ought to be a ground state.

\subsubsection{Common-Origin plot}
\label{COP}

As in Ref.~\onlinecite{SH}, we use the Common-Origin plot to easily see which directions the spins point in spin space. To construct a Common-Origin plot using a spin configuration determined by Iterative Minimization, we simply plot all the spin vectors with their tails at the origin. This diagnostic lets us easily see which directions the spins point in, but it doesn't give any information about how the spins are arranged on the lattice.

One can also make Common-Origin plots using only a subset of the entire spin configuration. For example, we might make a Common-Origin plot using only spins in one of the F.C.C. sublattices. To give another example, in cases where the lattice sites break up into sublattices according to site energy, we can make a Common-Origin plot using only spins with a certain site energy. Making Common-Origin plots of subsets of spins like these is often very helpful for finding a symmetrical basis for spin space to rotate the spin configuration into. It is much easier to think of possible idealized forms for a ground state when the numerical spin configuration obtained from Iterative Minimization has been rotated into the most symmetrical basis for spin space.

\subsection{Least Squares Fits and Variational Optimization}

\SAVE{MFL never did any variational optimization on the approximate parameterizations of the more complicated states. }

Many of the ground states we find in Iterative Minimization simulations are quite complicated, owing to the fact that they contain small contributions from many wavevectors besides the optimal Luttinger-Tisza wavevectors. In cases like these, it is desirable to obtain an approximate parameterization of the ground state. This approximate parameterization can then be used to obtain a more complete understanding of the state. To find these approximate parameterizations we perform a least squares fit of the numerical spin configuration in each F.C.C. sublattice to sines and cosines of the optimal Luttinger-Tisza wavevectors. If the spin configuration is rotated into a symmetrical basis in spin space, then this method usually yields approximate parameterizations which reveal the main structure of the ground state. 
\SAVE{Finding a good basis in spin space to rotate the spin configuration into is an interesting problem in itself. In practice, we usually find these good bases through a combination of analysis of the Common-Origin plot and trial and error.}

When the least squares fit is simple enough (i.e. it only involves one or two wavevectors), we can idealize the approximate parameterization into a guess for the closed-form parameterization of the ground state. Usually this guess involves a number of unknown parameters (e.g. wavevectors or cone-angles) whose values we compute by analytically solving for the values of these parameters which minimize the energy of the idealized state. 
This procedure is known as ``variational optimization"~\cite{SH}.

When we present parameterizations of spin configurations, we give a formula $\mb{S}_{\al}(\mb{r})$, which represents the spin configuration as a function of position in the $\al^{th}$ F.C.C. sublattice. We usually give the components of $\mb{S}_{\al}(\mb{r})$ in a Cartesian basis $\{ \xhat, \yhat, \zhat \}$ for spin space. Some states, however, take on their simplest form when written in terms of a basis for spin space which rotates as one moves from site to site in real space. 
It is important to remember that the basis vectors for spin space have no special orientation relative to the crystal lattice,
since equation \eqref{HeisHam} is invariant under a uniform rotation of all the spins.
\SAVE{We are free to choose the most convenient (most symmetrical) basis for spin space.}

\subsection{Projection of Stacked or Columnar States onto Lower Dimensional Lattices}
\label{sec:StacksAndColumns}

In this section we discuss a method for analyzing \emph{stacked} and \emph{columnar} states, in which the spin configuration is independent of one or two of the spatial coordinates, respectively, by projecting the lattice and interactions down onto an equivalent one or two-dimensional lattice.

A \emph{stacked} state is one in which there is a certain direction in real space, say $\hat{\mb{q}}$,  
such that within every plane normal to $\what{\mb{q}}$,  all the spins have the same direction.
Then we can project the three-dimensional lattice 
down onto an equivalent one-dimensional lattice 
(as elaborated in Ref.~\onlinecite{SH}, Section V B),
in which each spin represents an entire plane of spins 
from the original three-dimensional lattice. 
The three-dimensional spin interactions  are correspondingly
projected to the one-dimensional chain:
the interaction $j_{ij}$ between $\mb{s}_i$ and another  spin $\mb{s}_j$ in the chain,
is the sum of all interactions between some particular spin that projects to $\mb{s}_i$ 
and {\it any} spin in the plane that projects to $\mb{s}_j$. 

A \emph{columnar} state is a state in which there is a certain direction $\hat{\mb{n}}$, such that 
in any column of sites parallel to $\what{\mb{n}}$,
all the spins point in the same direction.
In this case, we can project the three-dimensional lattice and its interactions down onto an equivalent 
two-dimensional lattice,  in which each spin represents an entire line of spins in the original three-dimensional lattice. 
Similarly to the stacked case, the interaction between two spins $\mb{s}_i$ and $\mb{s}_j$ in this two-dimensional lattice is equal to the sum of all interactions in three dimensions between one spin in the preimage of $\mb{s}_i$ and all spins in
the preimage of $\mb{s}_j$.

If the result of an Iterative Minimization simulation is a \emph{stacked} or \emph{columnar} state, then we can make some progress towards understanding that state by projecting the lattice and interactions down onto a lower dimensional lattice. This lower dimensional problem is typically easier to analyze, and in cases where the lower dimensional lattice is a Bravais lattice, we can immediately use the Luttinger-Tisza method to rigorously justify the ground state. If a \emph{stacked} or \emph{columnar} state is the ground state of the three-dimensional lattice, then it must also be the ground state of the lower dimensional lattice (but the converse is not true).

The mapping of stacked states was introduced in 
Ref.~\cite{SH} for the study of conic spiral
states, which break up into a family of parallel
planes such that the spin direction is uniform within
each plane.
The state can be optimal in three dimensions only if
its projection is optimal on the one-dimensional lattice, but the
converse is not true.  (That is, a state might be
optimal on the one-dimensional lattice, but in 3D another state
that is not stacked in that way might have an even lower
energy.)

The one-dimensional lattice can have a non-coplanar ground state
only if it is non-Bravais lattice, i.e. only if
the mapped sites are inequivalent by translation.
Thus, it appears that the parallel layers of spins
in the three-dimensional lattice must be {\it unequal}.
This happens in the Octahedral lattice of Ref.~\cite{SH}
for (100) layers; in the Pyrochlore lattice
the $(111)$ stacking
consists of triangular-lattice linking layers, alternating with
kagom\'e-lattice layers that have three times as many spins.

\SAVE{
We now demonstrate this procedure with a simple example. Consider a state on the simple cubic lattice which is stacked along the $[001]$ direction and suppose that we include interactions between spins out to third nearest neighbors (so we have interactions $J_1$, $J_2$ and $J_3$). Denote the corresponding interactions in the one-dimensional lattice by $j_1$, $j_2$, $j_3$, etc. 
The first nearest neighbor interaction in the equivalent one-dimensional lattice must include all the interactions that a spin in a $(001)$ plane has with spins in the adjacent $(001)$ planes. A spin in the simple cubic lattice has six nearest neighbors but only one of these first nearest neighbor spins lie in each adjacent $(001)$ plane (the other four lie in the same plane as that spin). Each spin in the simple lattice also has 12 second nearest neighbors, four in each adjacent $(001)$ plane and four in its own plane. Finally, each spin in the simple cubic lattice has eight third nearest neighbors, four in each adjacent $(001)$ plane and zero in its own plane. So the corresponding first nearest neighbor interaction in the one-dimensional lattice is
$j_1 = J_1 + 4J_2 + 4J_3$.
Since the $J_1$, $J_2$ and $J_3$ interactions in the simple cubic lattice don't act between spins 
that are separated by more than a distance of 1 in the $(001)$ direction, the only interaction in 
the equivalent one-dimensional lattice is this first nearest neighbor interaction $j_1$. 
In this case, the one-dimensional lattice is a Bravais lattice, so one could immediately use 
the Luttinger-Tisza method to rigorously find a parameterization for the ground state.}

In the case of the Pyrochlore lattice, stacked states in which $\what{\mb{q}}$ is a $\langle100\rangle$ direction project 
onto a one-dimensional Bravais lattice; columnar states in which  $\what{\mb{n}}$ is a $\langle100\rangle$ direction project 
onto the two-dimensional non-Bravais lattice known as the Checkerboard lattice~\cite{MC98}.
States which are stacked along a $\langle111\rangle$ direction project down onto a 
one-dimensional non-Bravais lattice with a basis of two kinds of site,
which are projected respectively from  a triangular lattice layer of spins and a Kagome lattice layer of spins. 
The ground states of \eqref{HeisHam} on this two-site chain lattice, with various couplings,
were studied extensively in Ref.~\onlinecite{SH}.

\section{Overview of Results}

In this section we present a brief overview of our results. Most of our work was dedicated to characterizing the ground states of \eqref{HeisHam} for all possible combinations of first and second nearest neighbor interactions, culminating in the $J_1$-$J_2$ phase diagram. We also briefly explored a few ground states that include interactions beyond second nearest neighbors.

\subsection{The $J_1$-$J_2$ Phase Diagram}

\begin{figure}[t]
  \caption{The $J_1$-$J_2$ Phase Diagram for the Pyrochlore Lattice. It shows the regions of the $J_1$-$J_2$ phase diagram in which the Ferromagnetic (F.M.), Kawamura, $\mb{q}=\mb{0}$, $\left\{q00\right\}$ Planar Spiral, Double-Twist (D.T.), Multiply-Modulated Commensurate Spiral (M.M.C.S.) and Cuboctahedral Stack (C.S.) states are found. Solid lines represent first order phase transitions and dashed lines represent second order phase transitions. The dashed-dotted line along the negative $J_1$ axis indicates that there are a vast number of degenerate ground states along that line and that the Kawamura and $\mb{q}=\mb{0}$ states are members of that family of degenerate states. We call this a ``degenerate" phase transition.
}
\vskip 10pt
  \centering
    \includegraphics[width= .5\textwidth]{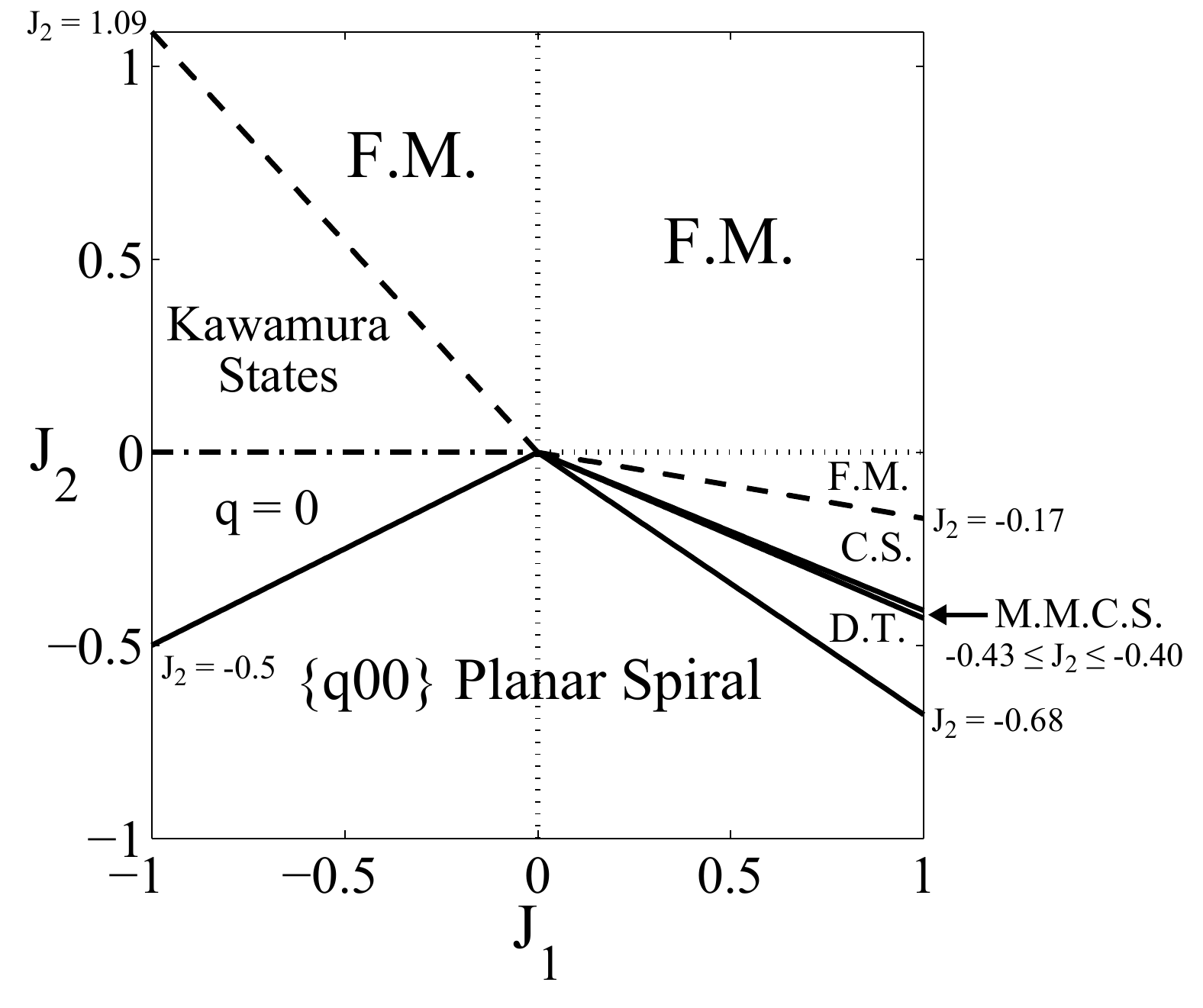}
\label{fig:PhaseDiag}
\end{figure}

The $J_1$-$J_2$ phase diagram is shown in figure \ref{fig:PhaseDiag}. The ground states that we find in this region of parameter space are summarized in table \ref{J1J2States}. The two largest regions in the phase diagram are the Ferromagnetic (F.M.) region and the $\left\{q00\right\}$ Planar Spiral region. In the region $J_1 > 0$, $J_2 > 0$ the ground state is ferromagnetic (section \ref{FM}). This is true even in the case of a pure ferromagnetic $J_1$ interaction or a pure ferromagnetic $J_2$ interaction. The reason for this is that the $J_1$ and $J_2$ interactions act between spins in different F.C.C. sublattices, so they couple together all of the spins in the lattice. The Ferromagnetic state extends into the $J_1<0$, $J_2>0$ and $J_1>0$, $J_2<0$ regions of the $J_1$-$J_2$ phase diagram, although it is quickly destabilized by the antiferromagnetic interactions found in those regions. 

\begin{table}[t]
	\begin{center}
	\renewcommand{\arraystretch}{1.2}
	  \begin{tabular}{ | c | c | c | c | c |}
	\hline
	 Ground State & $J_1$ &  $J_2$ & NCP? (y/n) &  $\QLT$ \\ \hline \hline
	Ferromagnetic (F.M.) & $+,-$ & $+,-$ & n & (0,0,0) \\ \hline
	$J_1 < 0$ & $-$ & $0$ & n & Any \\ \hline
	$\left\{q00\right\}$ Planar Spiral & $+,-$ & $-$ & n & $\left\{q00\right\}$ \\ \hline
	$\mb{q}=\mb{0}$ & $-$ & $-$ & n & (0,0,0) \\ \hline
	Cuboctahedral Stack (C.S.) & $+$ & $-$ & y & $\left\{\tfrac{1}{2}\tfrac{1}{2}\tfrac{1}{2}\right\}$\\ \hline
	Kawamura States & $-$ & $+$ & y & $\left\{\tfrac{3}{4}\tfrac{3}{4}0\right\}$ \\ \hline
	Double-Twist (D.T.) & $+$ & $-$ & y & $\left\{\tfrac{3}{4}\tfrac{3}{4}0\right\}$ \\ \hline
	Multiply-Modulated & $+$ & $-$ & y & $\left\{\tfrac{3}{4}\tfrac{1}{2}\tfrac{1}{4}\right\}$ \\ 
	Commensurate Spiral (M.M.C.S.) & &  & &  \\
	    \hline
	  \end{tabular}
	\caption{Summary of the ground states found in the $J_1$-$J_2$ phase diagram. It tells the region of the phase diagram where each state is found, whether or not the state is non-coplanar (NCP), and also gives the optimal Luttinger-Tisza wavevector $\QLT$ for each state.} 
	\label{J1J2States}
	\end{center}
\end{table}

The $\left\{q00\right\}$ Planar Spiral (section \ref{100spiral}) is a stacked state in which the spins spiral in a plane in spin space as one walks in a particular $\langle100\rangle$ direction in the lattice. This state exists in both the $J_1<0$, $J_2<0$ and $J_1>0$, $J_2<0$ regions of the $J_1$-$J_2$ phase diagram. In the special case where $J_1 = 0$ and $J_2<0$, the $\left\{q00\right\}$ Planar Spiral is a $120\dg$ spiral in a $\langle100\rangle$ direction in the lattice. This special case corresponds to the negative $J_2$ axis in figure \ref{fig:PhaseDiag}, which lies within the region where the $\left\{q00\right\}$ Planar Spiral is the ground state.

The next largest region in figure \ref{fig:PhaseDiag} is the region in the $J_1<0$, $J_2>0$ quadrant where the ground states are the three Kawamura States (section \ref{KawamuraStates}); the Sextuplet-q state, the Quadruplet-q state of type 1 and the Quadruplet-q state of type 2. The Fourier Transforms of these three states consist primarily of $\left\{\tfrac{3}{4}\tfrac{3}{4}0\right\}$ wavevectors. At $J_1 \approx -1.09J_1$ the Kawamura States transition to the ferromagnetic state.

In the upper part of the $J_1<0$, $J_2<0$ quadrant of figure \ref{fig:PhaseDiag} the ground state is the $\mb{q}=\mb{0}$ state (section \ref{qEquals0}), in which all spins in the same F.C.C. sublattice are parallel to each other, so that the spins point in only four directions in spin space. These four spin directions also add to zero because of the antiferromagnetic $J_1$ interaction . At $J_2 = 0.5J_1$ this state transitions to the $\left\{q00\right\}$ Planar Spiral state.

Finally, there are three more ground states in the $J_1>0$, $J_2<0$ quadrant of figure \ref{fig:PhaseDiag}. The first of these is the Cuboctahedral Stack (C.S.) state (section \ref{CS}), a non-coplanar state in which the spins point towards the eight corners and 12 edge-midpoints of a cube in spin space. As $J_2$ is made more negative there is a transition to the Multiply-Modulated Commensurate Spiral (M.M.C.S.) state (section \ref{MMCS}), a non-coplanar state in which the spins perform two different kinds of spirals which are controlled by two distinct wavevectors. Lastly there is the Double-Twist (D.T.) state (section \ref{DoubleTwist}), a non-coplanar state in which the spins perform two different kinds of spirals but those two kinds of spirals are controlled by the same kind of wavevector.

We have arranged our discussion of these states roughly in the order of least complicated to most complicated. In the section dealing with the simpler states (section \ref{SimpleStates}) we have grouped together states which are closely related. So, for example, the discussion of the $\mb{q}=\mb{0}$ state comes directly after the discussion of the ground states of the pure $J_1 < 0$ antiferromagnet, of which it is a special case. The Cuboctahedral Stack state is the most complicated state for which we still have an exact parameterization and energy per site, so our section on this state (section \ref{CS}) can be said to bridge the gap between the less complicated and more complicated states.  

\subsection{States with $J_3$ and $J_3'$}

In our exploration of states involving the third nearest neighbor interactions $J_3$ and $J_3'$, we again found the Cuboctahedral Stack state, and we also found a new kind of Alternating Conic Spiral state. We found that one can also stabilize the Cuboctahedral Stack state with a ferromagnetic $J_1$ interaction, a ferromagnetic $J_3$ interaction and an antiferromagnetic $J_3'$ interaction. We conjecture that the two regions of parameter space in which the Cuboctahedral Stack is the ground state (i.e. this region and the region in the $J_1$-$J_2$ plane) are smoothly connected.

When the magnitude of the antiferromagnetic $J_3'$ interaction becomes much larger than the magnitudes of the ferromagnetic $J_1$ and $J_3$ interactions, we find a new kind of Alternating Conic Spiral state (appendix \ref{ACS}). This state is different from the alternating conic spirals discovered in Ref.~\onlinecite{SH} in the sense that the wavevector which controls the spiraling behavior and the wavevector which controls the alternating behavior are not parallel to each other. 

\section{Simple States}
\label{SimpleStates}

We start our tour of the phase diagram with the most elementary
cases, in that there is only one coupling, and/or there are
ground states with the periodicity of the unit cell.
All these states are exact Luttinger-Tisza states whose exact energies are known,
and in none of these cases is a non-coplanar state forced.

\subsection{Ferromagnetic State}
\label{FM}

The simplest state we find (and the simplest possible ground state) is the ferromagnetic state, in which every spin points in the same direction. This state is obviously the ground state in the region where $J_1$ and $J_2$ are both greater than zero, but we also find it in the region $J_1 < 0$, $J_2 > 0$ when $J_2 \geq -1.09J_1$ and in the region $J_1 > 0$, $J_2 < 0$ when $J_2 \geq \left(-\tfrac{3}{8}+\tfrac{\sqrt{6}}{12}\right)J_1$. The energy per site of the ferromagnetic state is 
	\beq
		\Eavg = -3J_1 - 6J_2 \label{Eferro}\ .
	\eeq 
If the interactions are $J_1$, $J_3$ and $J_3'$ (section \ref{sec:CSJ3}), the energy per site of the ferromagnetic state is 
	\beq
		\Eavg = -3J_1 - 3J_3 - 3J_3' \label{EferroJ3}\ .
	\eeq 

We use these expressions for the energy per site of the ferromagnetic state to predict where this state transitions to competing states in the regions $J_1 > 0$, $J_2 < 0$ and $J_1 < 0$, $J_2 > 0$, and also in regions of the phase diagram where $J_3$ and $J_3'$ are non-zero.

\subsection{Pure $J_1<0$}
\label{J1}

The ground state of \eqref{HeisHam} in the case of a pure antiferromagnetic first nearest neighbor interaction $J_1 < 0$ is well known to be any state in which the sum of the four spins on each tetrahedron is equal to zero. 
One can prove this using the Cluster method~\cite{Ly64}, in which one finds a way to rewrite 
the Hamiltonian as a sum of disjoint terms:
if a state can be exhibited in which each term is separately minimized,
this must be a ground state (and all other ground states must have the same property)

Let the tetrahedra in the lattice be indexed by the Greek letter $\eta$ and let the spin in F.C.C. sublattice $\al$ on the $\eta^{th}$ tetrahedron be $\mb{s}_{\eta,\al}$. 
Then the pure-$J_1$ Hamiltonian takes the form
	\beq 
		\Ham = -\frac{1}{2}J_1\sum_{\eta} \left|\mb{L}_{\eta}\right|^2+ constant\ , 
        \label{HamJ1}
	\eeq 
where $\mb{L}_{\eta} = \sum_{\al= 0}^{3} \mb{s}_{\eta,\al}$, i.e. 
the sum of the four spins on the $\eta^{th}$ tetrahedron. 

Eq.~\eqref{HamJ1} is a sum over disjoint terms, 
each of which is minimized by 
        \beq
              \mb{L}_{\eta} = \mb{0}\ .
        \label{eq:tetzero}
        \eeq 
Furthermore it is easy to find an example configuration
that satisfies condition \eqref{eq:tetzero} simultaneously 
on all tetrahedra, 
so we can rigorously conclude this condition is true for all $\eta$ in {\it all} ground states.
In any such a state, the site energy is $\Esite = J_1$ for {\it every} spin,
so the energy per site is also
	\beq
		\Eavg = J_1\ .
	\label{eq:J1energy}
	\eeq
These states have an extensive degeneracy,
meaning the number of parameters needed to specify the spin configuration
is proportional to the number of spins in the system.
The $\mb{q} = \mb{0}$ states of Sec.~\ref{qEquals0} and the Kawamura states of 
Sec.~\ref{KawamuraStates} still satisfy the constraint \eqref{eq:tetzero} so they are
specific subsets of the ground states for the pure $J_1 < 0$ case.

\subsection{$\mb{q} = \mb{0}$ State(s) ($J_1 < 0$ and $J_2 < 0$)}
\label{qEquals0}

In the region $J_1 < 0$, $\tfrac{1}{2}J_1 \leq J_2 \leq 0$ we find a state in which all spins in the same F.C.C. sublattice are parallel and the spins from the four F.C.C. sublattices all sum to zero. To be more precise, we have $\mb{S}_{\al} = \mb{n}_{\al}$, where the $\mb{n}_{\al}$ are all constant unit vectors which satisfy
	\beq
		\sum_{\al=0}^3\mb{n}_{\al} = \mb{0} \ . 
                \label{q0constraint}
	\eeq

We call this state a $\mb{q} = \mb{0}$ state because the wavevector characterizing the spin configuration in each F.C.C. sublattice is the zero wavevector. The site energy for each spin in this state is $\Esite = J_1 + 2J_2$ so this state has an energy per site equal to 
	\beq
		\Eavg = J_1 + 2J_2\ . \label{q0H}
	\eeq

For sufficiently small $J_2$, at least, the ground state must be a subset of the
pure-$J_1$ ground state manifold of Section~\ref{J1}, 
selected by $J_2$ as a degenerate perturbation.
This state is most quickly understood by using the equivalence of a small antiferromagnetic 
(ferromagnetic) $J_2$ interaction with a small ferromagnetic (antiferromagnetic) $J_3$ interaction,
within that degenerate manifold (see Ref.~\onlinecite{Che08} for a proof of this fact).  The reason for this equivalence is the following.
The $J_2$ and $J_3$ interactions of spin $i$ can be gathered into contributions from six 
second-neighbor  tetrahedra, each of which includes a different $J_1$ neighbor of spin $i$.
But in view of \eqref{eq:tetzero}
the sum of the two $J_2$ neighbors, the one $J_3$ neighbor, and the one $J_1$ neighbor
in each of those tetrahedra is zero, hence the sum of the $J_1$, $J_2$, and $J_3$ energies 
is also zero -- but the total $J_1$ energy is a fixed constant within the degenerate
manifold satisfying \eqref{eq:J1energy}.

So, we can trade this problem for that of a small $J_3>0$.
Since the $J_3$ bonds connect sites of same F.C.C. sublattice, 
the $J_3$ term is optimized when all spins in the same F.C.C. sublattice 
are parallel (so $\mb{q}=\mb{0}$ by definition).  This is compatible with
\eqref{eq:tetzero}: the necessary and sufficient condition is 
Eq.~\eqref{q0constraint}, as was claimed.

Thus the $\mb{q} = \mb{0}$ states form a continuous two-dimensional manifold
of degenerate states, inequivalent by rotational symmetry,
and parametrized by two angles, $\theta$ and $\phi$.   
We can define $\theta \in [0,\pi]$ to be the angle between $\mb{n}_0$ and $\mb{n}_1$ --
this must also be the angle between $\mb{n}_2$ and $\mb{n}_3$, since
$|\mb{n}_0+\mb{n}_1|=|\mb{n}_2+\mb{n}_3|$; then we can take $\phi\in[0,\pi]$
to be the angle between the plane of $(\mb{n}_0, \mb{n}_1)$
and that of $(\mb{n}_2, \mb{n}_3)$.

\SAVE{Is $\phi\in[0,\pi]$ or $[0,\pi/2]$?
MFL thinks it is $\pi$, since if you start with 
$\mb{n}_2= -\mb{n}_0$ and $\mb{n}_3= -\mb{n}_1$ and rotate by $\pi$ 
you swap and get $\mb{n}_3= -\mb{n}_0$ and $\mb{n}_2= -\mb{n}_1$. 
If you rotate by any more than $\pi$ you get a state which is equivalent by relfection through the ($\mb{n}_0, \mb{n}_1$) plane to a state you already got with $\phi < \pi$.}

\SAVE{Note $\theta$, $\phi \in [0, \pi]$ since $\theta$ is the smaller of the two angles between $\mb{n}_0$ and $\mb{n}_1$, and since $\phi$ and $2\pi-\phi$ produce the same state because of the reflection symmetry through the plane spanned by $\mb{n}_0$ and $\mb{n}_1$. }

Because the $\mb{q}=\mb{0}$ family satisfies the constraint \eqref{q0constraint}, 
this state is also a ground state of \eqref{HeisHam} in the case of a pure first nearest neighbor interaction $J_1 < 0$. Evidently, the set of $\mb{q} = \mb{0}$ states is just a subset of the ground states for the pure $J_1 < 0$ interaction, and these states are selected out by the antiferromagnetic $J_2$ interaction.

\begin{figure}[t]
  \caption{The four eigenvalues $\lam_{\nu}(\mb{q})$ of the matrix $\un{\JLT}(\mb{q})$ plotted along the $\Gamma$ $\to$ $X$ $\to$ $W$ $\to$ $L$ $\to$ $\Gamma$ $\to$ $K$ $\to$ $X$ route through the the Brillouin zone of the F.C.C. lattice for $J_1 = -1$, $J_2 = -0.25$, where the ground state is the $\mb{q}=\mb{0}$ state. Recall that $\Gamma = (0,0,0)$, $X = 2\pi(1,0,0)$, $W = 2\pi(1,\tfrac{1}{2},0)$, $K = 2\pi(\tfrac{3}{4},\tfrac{3}{4},0)$ and $L = 2\pi(\tfrac{1}{2},\tfrac{1}{2},\tfrac{1}{2})$. 
 The largest eigenvalue of $\un{\JLT}(\mb{q})$ takes on its maximum value at the $\Gamma$ point $\mb{q} = \mb{0}$, confirming that the optimal L.T. wavevector in this region is $\QLT = \mb{0}$, as expected.}
\vskip 10pt
  \centering
    \includegraphics[width= .5\textwidth]{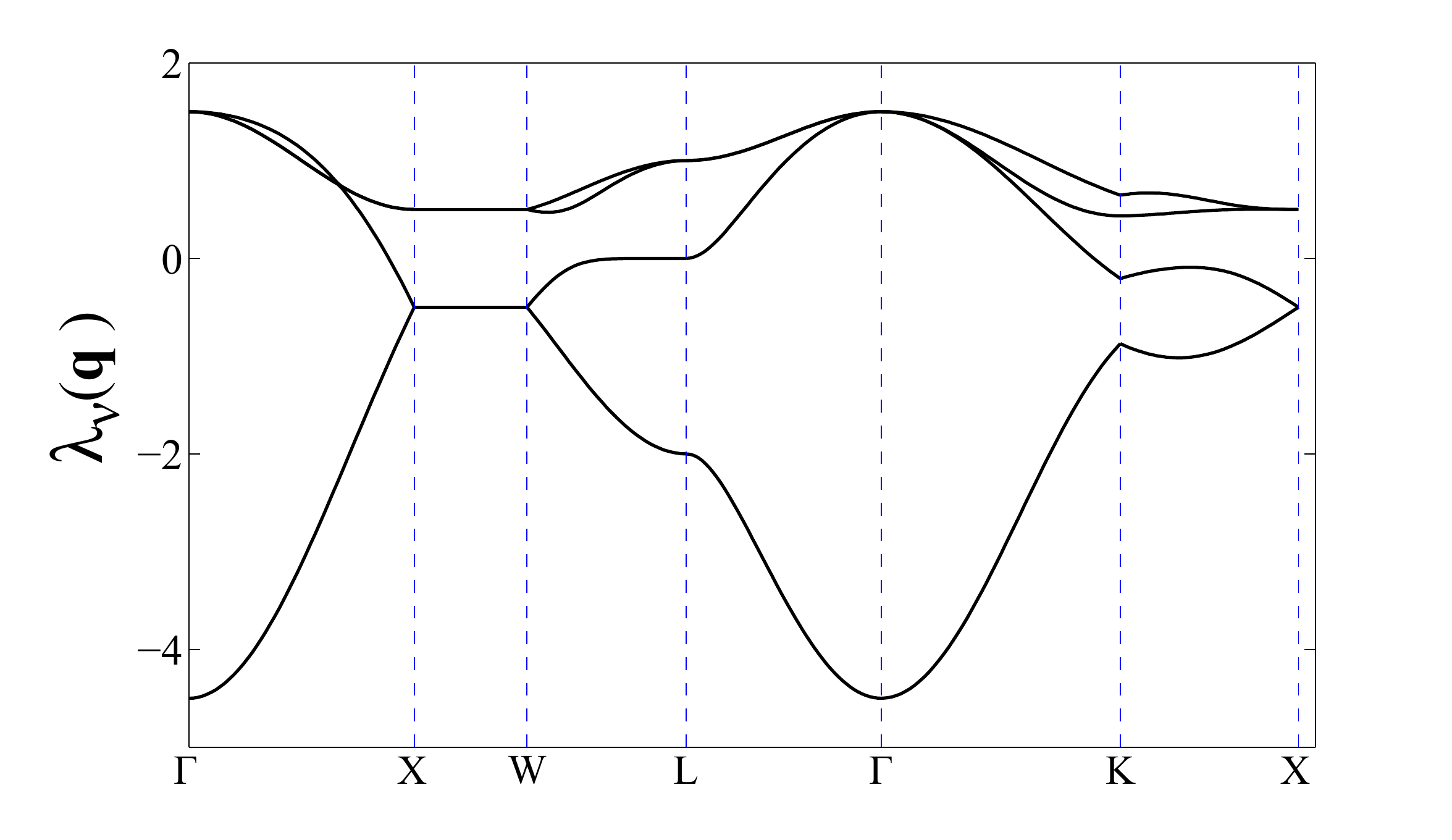}
\label{fig:BandsQ0}
\end{figure}

Figure \ref{fig:BandsQ0} shows the four eigenvalues $\lam_{\nu}(\mb{q})$ of the matrix $\un{\JLT}(\mb{q})$ plotted along the $\Gamma$ $\to$ $X$ $\to$ $W$ $\to$ $L$ $\to$ $\Gamma$ $\to$ $K$ $\to$ $X$ route through the Brillouin zone of the F.C.C. lattice for the parameter values $J_1 = -1$, $J_2= -0.25$. 
We see that the largest eigenvalue has its maximum at the $\Gamma$ point 
$\mb{q} = \mb{0}$, so the optimal L.T. wavevector is $\QLT=\mb{0}$, not only for small $J_2$
but in a considerable interval.  Since we constructed normalized spin states using just
these modes, they are rigorously the ground states, and Iterative Minimization simulations
confirm this.

\subsection{Pure $J_2<0$}
\label{J2spiral}

Unlike the case of a pure $J_3$ interaction, in which the Hamiltonian breaks down into uncoupled sublattices, the pure $J_2$ interaction acts between sublattices, so that every spin is coupled to every other one. In the case of a pure antiferromagnetic second nearest neighbor interaction, we find that the ground state is a $120^{\circ}$ planar spiral stacked in a $\langle100\rangle$ direction in the lattice. We can get an understanding of this state by \textit{assuming} (based on evidence from Iterative Minimization simulations) that the state is stacked along a $\langle100\rangle$ direction and then projecting the interactions down onto an equivalent one-dimensional lattice. We can then apply the Luttinger-Tisza method~\cite{LT,Kapl06} to this equivalent one-dimensional lattice. 
We use the method outlined in subsection \ref{sec:StacksAndColumns} to determine the effective interactions in the equivalent one-dimensional lattice. 

Let us take the stacking to be along the $z$-direction.
The $(001)$ planes of sites with the same spin direction are separated by a distance of $\tfrac{1}{4}$. 
A spin has four of its second nearest neighbors in each adjacent plane 
(separated by a distance $\tfrac{1}{4}$ in the z direction) 
and two more second nearest neighbors in planes that are two steps away 
(i.e. separated by $\tfrac{1}{2}$ in the z direction). 
Thus, the equivalent one-dimensional lattice has both first and second neighbor interactions, 
$j_1 = 4J_2$ and $j_2 = 2J_2 = \tfrac{1}{2}j_1$. In addition, the one-dimensional lattice is a Bravais lattice, because every site in the same constant-z plane has the same number of second nearest neighbors in the planes above and below it.

The optimal wavevector takes the form $\mb{q} = (0,0,q)$ in the three-dimensional lattice. In the one-dimensional lattice
the state will be characterized by a wavenumber $q_{1d}$. We have $q = q_{1d}/2$ since in the three-dimensional lattice the minimum separation in the z direction between two spins in the same F.C.C. sublattice with the same x and y coordinates is $\tfrac{1}{2}$, whereas the minimum separation between two spins in the one-dimensional Bravais sublattice is $\tfrac{1}{4}$. Since the minimum separation in the z direction between two spins in the same sublattice is doubled when going from the one-dimensional lattice to the three-dimensional lattice, the wavevector is halved. 

Applying the Luttinger-Tisza method to the one-dimensional chain gives an expression for $\tilde{J}(q_{1d})$, the Fourier transform of the couplings in the one-dimensional chain:
	\beq
		\tilde{J}(q_{1d}) = 4J_2\left(2\cos\left(\frac{q_{1d}}{4}\right) + \cos\left(\frac{q_{1d}}{2}\right)\right)
	\eeq
Maximizing this we find that the optimal (one-dimensional) wavenumber is given by 
	\beq
		\cos\left(\frac{q_{1d}}{4}\right) = - \frac{1}{2} 
	\eeq
so we have $q_{1d} = \tfrac{8\pi}{3}$, which means that spins separated by a distance of $\tfrac{1}{4}$ in the z-direction (first nearest neighbors in the one-dimensional lattice) are rotated $120^{\circ}$ from each other. So the ground state is a $120\dg$ spiral in the z-direction. This state has the form
	\beq
		\mb{S}_{1d}(z) = \cos\left(\frac{8\pi}{3}z\right)\xhat + \sin\left(\frac{8\pi}{3}z\right)\yhat  
	\eeq
in the one-dimensional lattice, up to an arbitrary (constant) phase angle. In the three-dimensional lattice we have $q =
\tfrac{4\pi}{3}$, and we can parameterize this same state as
\begin{subequations}
	\beqa
		\mb{S}_{\al= 0,3}(\mb{r}) &=& \cos\left(\frac{4\pi}{3}z\right)\xhat - \sin\left(\frac{4\pi}{3}z\right)\yhat \label{J2spiral1}\\
		 \mb{S}_{\al= 1,2}(\mb{r}) &=& \cos\left(\frac{4\pi}{3}z + \frac{\pi}{3}\right)\xhat \nonumber \\
 &  & \ \ \ \ \ \ \ \ \ \ \ \ \  + \ \ \sin\left(\frac{4\pi}{3}z + \frac{\pi}{3}\right)\yhat \label{J2spiral2}
	\eeqa
\end{subequations}
in the separate F.C.C. sublattices.

The site energy of each spin is given by $\Esite = 3J_2$ so this state has an energy per site
	\beq
		\Eavg = 3J_2 \ . \label{J2H}	
	\eeq

\SAVE{We see that the $120^{\circ}$ spiral arises naturally in the context of the one-dimensional Bravais lattice as a consequence of the fact that $j_2 = \tfrac{1}{2}j_1$ and both $j_1$ and $j_2$ are antiferromagnetic interactions (they are both negative).}

The $120^{\circ}$ spiral 
is just a special case of the more general $\left\{q00\right\}$ planar spiral state,
discussed in the next subsection (\ref{100spiral}),
which involves both $J_1$ and $J_2$ interactions.

\subsection{$\left\{q00\right\}$ Planar Spiral ($J_2 < 0$ and $J_1 > 0$ or $J_1 < 0$)}
\label{100spiral}

In the region $J_1 < 0$, $J_2 \leq \tfrac{1}{2}J_1$ and also in the region 
$J_1 > 0$, $J_2 \leq -0.68J_1$, we find a planar spiral state which is stacked in a 
$\langle100\rangle$ direction, which (as in Sec.~\ref{J2spiral}) we again
take along $z$.
\SAVE{Thus this state is characterized by a wavevector $\mb{q} = (0,0,q)$.}
This generalized spiral can also be understood by mapping the interactions down onto 
an equivalent one-dimensional lattice. 
now with couplings $j_1 = 2J_1 + 4J_2$ and $j_2 = 2J_2$. 
\SAVE{(Because a spin in a constant-z plane has two of its first nearest 
neighbors in each constant-z plane adjacent to it, 
and because the second nearest neighbors of each spin are distributed 
in the way described in subsection \ref{J2spiral} on the pure $J_2 < 0$ spiral.)}
Applying the Luttinger-Tisza method to this one-dimensional lattice gives 
	\beq
		\tilde{J}(q_{1d}) = 2j_1\cos\left(\frac{q_{1d}}{4}\right) + 2j_2\cos\left(\frac{q_{1d}}{2}\right) \ .
	\eeq
so the optimal one-dimensional wavenumber is given by
	\beq
		\cos\left(\frac{q_{1d}}{4}\right) = -\frac{j_1}{4j_2}
		 = -\frac{J_1 + 2J_2}{4J_2} \label{Qspiral}
	\eeq
in terms of the interactions in the three-dimensional lattice. The parameterization of this state in the three-dimensional lattice has the same form as equations \eqref{J2spiral1} and \eqref{J2spiral2}, but with $\tfrac{4\pi}{3}$ replaced by the wavenumber $q$ determined from \eqref{Qspiral} and the relation $q = q_{1d}/2$.
\SAVE{(So spins separated by a distance $\tfrac{1}{4}$ in the z-direction 
will be rotated from each other by $q$ radians.)}
The site energy for each spin in this state is
	\beq
		\Esite = \frac{J_1^2}{4J_2} + 3J_2\ ,
	\eeq
so $\Eavg = \Esite$.
In the special case of $J_1 = 0$, 
we recover the $120\dg$ spiral found in the previous section. 

Finally, we note that in order for equation \eqref{Qspiral} to have a solution for $q_{1d}$, we must have $-1 \leq -\tfrac{J_1 + 2J_2}{4J_2} \leq 1$. Therefore, this state can only exist in the region $J_1 < 0$ when $J_2 \leq \tfrac{1}{2}J_1$, and it can only exist in the region $J_1 > 0$ when $J_2 \leq -\tfrac{1}{6}J_1$.

\section{Cuboctahedral Stack}
\label{CS}

\LATER{8 CLH should fix up the intro here some time.}

In the region $J_1 > 0$, $J_2 < 0$ we find a state that we have named the Cuboctahedral Stack because the spins in each Kagome lattice layer perpendicular to a certain $\langle111\rangle$ direction in real space are arranged in the Cuboctahedral state found on the Kagome lattice in Ref.~\onlinecite{Do05} and on the Octahedral lattice in Ref.~\onlinecite{SH}. In each of those states the spins point towards the 12 vertices of a cuboctahedron or, equivalently, the 12 edge-midpoints of a cube. If we choose a basis in spin space in which the sides of this cube are perpendicular to $\langle100\rangle$ directions, the spins point in the twelve $\langle110\rangle$ directions in spin space.

\subsection{Structure of the Cuboctahedral Stack}

In the Cuboctahedral Stack state there is a distinguished direction in real space, which is one of the four $\langle111\rangle$ directions (we choose $[111]$ for the parameterizations below).
In the Kagome lattice layers stacked perpendicular to this direction, the spins point towards the 12 edge-midpoints of a cube. In the triangular lattice layers between the Kagome layers, the spins point towards the eight corners of that same cube.

This state is built out of three of the four $\left\{\tfrac{1}{2}\tfrac{1}{2}\tfrac{1}{2}\right\}$ type wavevectors and the fourth unused wavevector points in the stacking direction of the Kagome and triangular lattice layers; it can be written as
\begin{subequations}
	\beqa
	    \mb{S}_0 &=& \frac{1}{\sqrt{3}}\sum_{k=1}^{3}\cos(\mb{q}_k\cdot\mb{r})\ \hat{\mb{e}}_k 
         \label{CS0} \\
	    \mb{S}_{\al=1,2,3} &=& \frac{1}{\sqrt{2}}\sum_{k=1,\ k \neq \al}^{3}\cos(\mb{q}_k\cdot\mb{r})\ \hat{\mb{e}}_k ,
         \label{CS123}
	\eeqa
\end{subequations}
where the three wavevectors used in this parameterization are $\mb{q}_1 = 2\pi(-\tfrac{1}{2}, \tfrac{1}{2}, \tfrac{1}{2})$, $\mb{q}_2 = 2\pi(\tfrac{1}{2}, -\tfrac{1}{2}, \tfrac{1}{2})$ and $\mb{q}_3 = 2\pi(\tfrac{1}{2}, \tfrac{1}{2}, -\tfrac{1}{2})$.

\SAVE{(MFL:) Earlier versions (including my thesis) 
were missing a minus sign for the site energy of the 0th FCC sublattice.}
In this parameterization, the triangular layers consist of the spins in F.C.C. sublattice 0,  whose configuration is given by \eqref{CS0}, and the Kagome layers consist of the spins in F.C.C. sublattices 1, 2 and 3, whose configurations are given by \eqref{CS123}. The site energies for spins in the four F.C.C. sublattices are
\begin{subequations}
	\beqa
		\Esite^{(0)}&=& -\sqrt{6}J_1 \\
		\Esite^{(1,2,3)} &=& -J_1\left(1+\frac{\sqrt{6}}{3}\right)\ ,
	\eeqa
\end{subequations}
where the superscript indexes the F.C.C. sublattices. The site energy of a spin in a triangular layer is lower than the site energy of a spin in a Kagome layer. 
\subsection{Energy per spin in the cuboctahedral stack}

The energy per site in this state is then

	\beq
		\Eavg = -J_1\left(\frac{3}{4} + \frac{\sqrt{6}}{2}\right)\ . 
        \label{Estack}
	\eeq
Evidently, the $J_2$ contribution to the energy of this state completely cancels out, so that the total energy depends only on $J_1$. 

In fact, the lower bound on the energy of this state (from equation \eqref{Ebound}) is $-2J_1$, so even though the Cuboctahedral Stack is constructed solely from the optimal L.T. wavevectors, it does not achieve the lower bound on the energy. This is related to the fact that only three out of a total of four available symmetry related $\left\{\tfrac{1}{2}\tfrac{1}{2}\tfrac{1}{2}\right\}$ modes are used to construct this state.

\begin{figure}[t]
  \caption{The four eigenvalues $\lam_{\nu}(\mb{q})$ of the matrix $\un{\JLT}(\mb{q})$ plotted along the $\Gamma$ $\to$ $X$ $\to$ $W$ $\to$ $L$ $\to$ $\Gamma$ $\to$ $K$ $\to$ $X$ route through the Brillouin zone of the F.C.C. lattice for $J_1 = 1$, $J_2 = -0.2$, where the Cuboctahedral Stack (C.S) state is the ground state. Recall that $\Gamma = (0,0,0)$, $X = 2\pi(1,0,0)$, $W = 2\pi(1,\tfrac{1}{2},0)$, $K = 2\pi(\tfrac{3}{4},\tfrac{3}{4},0)$ and $L = 2\pi(\tfrac{1}{2},\tfrac{1}{2},\tfrac{1}{2})$. The largest eigenvalue of 
$\un{\JLT}(\mb{q})$ takes on its maximum value at the $L$ point, $\mb{q} = 2\pi(\tfrac{1}{2},\tfrac{1}{2},\tfrac{1}{2})$, the wavevector which characterizes the Cubocatahedral Stack state.}
\vskip 10pt
  \centering
    \includegraphics[width= .5\textwidth]{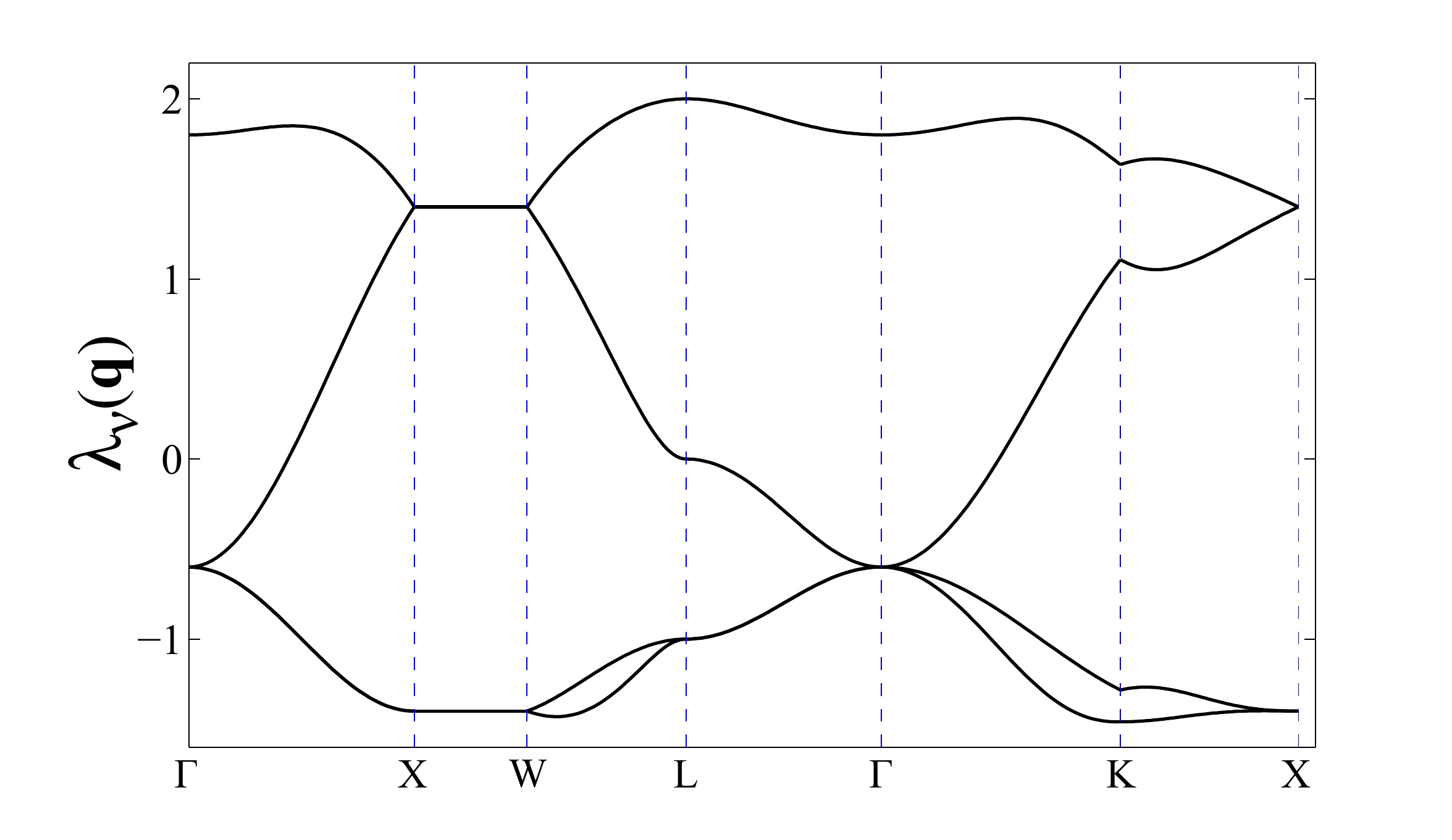}
\label{fig:BandsCS}
\end{figure}

Figure \ref{fig:BandsCS} shows the four eigenvalues $\lam_{\nu}(\mb{q})$ of the matrix $\un{\JLT}(\mb{q})$ plotted along the $\Gamma$ $\to$ $X$ $\to$ $W$ $\to$ $L$ $\to$ $\Gamma$ $\to$ $K$ $\to$ $X$ route through the Brillouin zone of the F.C.C. lattice for the parameter values $J_1 = 1$, $J_2= -0.2$. We see that the largest eigenvalue has its maximum at the $L$ point $\mb{q} = 2\pi(\tfrac{1}{2},\tfrac{1}{2},\tfrac{1}{2})$, confirming that the optimal L.T. wavevector is $\QLT = 2\pi(\tfrac{1}{2},\tfrac{1}{2},\tfrac{1}{2})$, as expected from Iterative Minimization simulations.

\subsubsection{Why energy is independent of $J_2$}

\LATER{8a CLH thinks this subsection goes a bit long, on rather technical things.  
Yet it is a noteworthy point.  It just occurred to me that maybe it's
appropriate to move this to a short appendix.  Also, if you insert some
answers to my questions about the LT viewpoint, it might end up in the main
text as 1 paragraph about why independent of J2 (citing the appendix
if needed) + 1 paragraph about LT.}

To see why the total energy \eqref{Estack} is independent of $J_2$,
we need to think more about the structure of this state. Of the four spins in every tetrahedron, three of them lie in a Kagome lattice layer and the fourth lies in a triangular lattice layer. Recall that every spin in a triangular lattice layer points in a $\langle111\rangle$ direction and every spin in a Kagome lattice layer points in a $\langle110\rangle$ direction. To approximately satisfy the ferromagnetic first nearest neighbor interaction, the three spins on a tetrahedron in a Kagome lattice layer will point in three $\langle110\rangle$ directions which surround the same corner of a cube in spin space. For example, these spins might point in the $[110]$, $[101]$ and $[011]$ directions. Then the dot product between any two of these three spins will be $\tfrac{1}{2}$. The fourth spin on this tetrahedron, the one which lies in a triangular lattice layer, will then point in the $[111]$ direction, the average of the directions the other three spins point in. The dot product of this fourth spin with any of the other three will then be $\tfrac{2}{\sqrt{6}} \approx 0.82$, which also favors the ferromagnetic first nearest neighbor interaction.

\LATER{9 My own big concern in this section is not the dot products in real space.
It is the LT picture.  And this is related to a reply I made somewhere above
(why you can't make a state with 4 wavevectors).  I also talked about it in
the conclusion of Sophia's paper, Sec VIII B 2 and VIII C.  If we took all four
of the (1/2 1/2 1/2) wavevectors, and let each have an orthogonal spin component,
we would certainly make a normalized state.  In the norm of site $i$, a given spin 
component associated with mode $\mu$ contributes proportionally to 
$|u^\mu_i|^2$, where $u^\mu_i$ is mode $\mu$ on site $i$.  
A given mode is NOT uniform on the sites, but if we add the squares of all
the symmetry-related modes, we must (by symmetry) get something that IS uniform.
In the present case, however, you are missing one mode, so how can the result
be uniform?  My best guess is that the spin directions associated with 
the three respective $q$ vectors are NOT orthogonal to each other, and the
cross-terms in the spin norm due to this, are somehow compensating for the
non-uniformity due to omitting one mode.}

\LATER{10 Fold in these paragraphs (related to LT picture of cuboc stack):
In Ref.~\onlinecite{SH}, Section III B 1, 
it was considered from the LT viewpoint~
how the cuboctahedral state might be generalized
to the kagom\'e lattice.  The implicit (and 
incorrect) assumption was that the spin configuration
should have all the point group symmetries of the lattice.
Thus, one would first identify a star of specially symmetric wavevectors
$\QQ_{1,2,3}$ with threefold multiplicity: in a face-centered
cubic structure, that can only be $\{1,0,0\}$ wavevectors.
Assume there are couplings $\{J_i\}$ for which
the LT optimal modes are on these wavevectors.
Naively, one would construct a linear combination of these,
using a different cartesian direction
for each mode, and obtain a non-collinear state.
However, an important caveat was shown there:
since the planes of sites defined by the $\{1,0,0\}$ wavevector are
all equivalent, the symmetry-related optimal modes are simply ``decoupled'':
the whole spin structure breaks up into sublattices
which can be rotated independently, with the inter-sublattice couplings
canceling.  We do not get a necessarily cuboctahedral phase.
\par
In the state we actually found, the wavevectors $\QQ_{1,2,3}$ come from
a star of fourfold multiplicity, out of which  one is unused.
We encounter a different worry in this case: the state lacks the full 
lattice symmetry, indeed the kagom\'e and trianglular layer spins are
inequivalent, so it is not generic for them to be normalized the same.
This appears to happen accidentally.
}  

\LATER{11 MFL: I think we need to save this for later. 
I added after the site energy equation (5.2c) that the lower bound on the energy in this region is actually $-2J_1$, which is a little bit lower than the energy of the C.S. state. So even though the C.S. is constructed solely from optimal wavevectors, it does not achieve the lower bound on the energy. I think this could be related to what you are saying about how we are using only 3 out of 4 symmetry equivalent modes to make this state and the fact that there is some non-uniformity.}

Because of this structure, the dot product between any spin in a triangular lattice layer (i.e. in F.C.C. sublattice 0) and any of its second nearest neighbors is always zero. The reason for this is that any second nearest neighbor of a spin in a triangular lattice layer will be in a Kagome lattice layer, so it will point in a $\langle110\rangle$ direction. Moreover, the $\langle110\rangle$ direction it points in will be one which surrounds a corner of the cube in spin space which is adjacent to the corner that the spin in the triangular lattice layer points towards, but this $\langle110\rangle$ direction will not point towards the midpoint of the edge that connects these two corners of the cube in spin space. It follows from this that the dot products of a spin in a triangular layer with its second nearest neighbors are all zero.

 The dot products between any spin in a Kagome lattice layer and its 12 second nearest neighbors are not all zero but they do add to zero. Of these 12 dot products, four of them are 0 (the dot products with spins in the adjacent triangular layers), four of them are $-\tfrac{1}{2}$ (the dot products with spins in the same Kagome layer) and four of them are $\frac{1}{2}$ (the dot products with spins in the neighboring Kagome layers which are beyond the adjacent triangular layers). 

\subsection{Cuboctahedral stack in the $(J_1,J_2)$ phase diagram}

As explained in the last subsection, when we fix $J_1$ and vary $J_2$, 
the energy of the Cuboctahedral Stack state is unchanging,
but the energy of any competing states (whose energy generally does 
depend on $J_2$) may cross lower than that of the Cuboctahedral Stack,
at which point  there will be a phase transition. 
In particular, comparing Eq.~\eqref{Estack} to
Eq.~\eqref{Eferro}, we find that the transition from the ferromagnetic state to the Cuboctahedral Stack should occur at 
	\beq
		J_2= \left(-\frac{3}{8}+\frac{\sqrt{6}}{12}\right)J_1 \approx -0.171J_1 \ , 
	\eeq
and this result is confirmed by simulations. Simulations also show that the Cuboctahedral Stack transitions to the Multiply-Modulated Commensurate Spiral state (section \ref{MMCS}) near $J_2 = -0.40J_1$.

\subsubsection{Transition between the Cuboctahedral Stack and ferromagnetic states}

\SAVE{MFL is not sure about the conclusion I made that the transition is continuous, 
but I think it seems most likely in light of the evidence. On the C.S. side of the transition I always find the C.S. state or an incompletely relaxed state that has a higher energy than the C.S. state. On the F.M. side, there really is a state near the transition that does better than either the C.S. or the F.M. state, and this intermediate state has characteristics of both states. As you get closer to the boundary this intermediate state looks like it is unfolding outwards from a conic axis from the F.M. state to the C.S. state.}

Because there are no variable parameters in the parameterization of the Cuboctahedral Stack state, we might expect that the transition between this state and the ferromagnetic state would be of first order. The results of Iterative Minimization simulations show that this is not the case. On the ferromagnetic side of the phase boundary we find a ``mixed" state near the transition point whose energy is lower than the energies of both the ferromagnetic and Cuboctahedral Stack states. The Fourier Transform of this mixed state shows peaks at $\mb{q}=\mb{0}$ as well as at the $\left\{\tfrac{1}{2}\tfrac{1}{2}\tfrac{1}{2}\right\}$ wavevectors that make up the Cuboctahedral Stack. In spin space the spins are arranged around a cone so that this state does have a net magnetic moment which points along the axis of that cone. 

If one looks in detail at the spin directions in the Common-Origin plot, one sees that the spins unfold from the conic axis in four sets of three spins and four sets of two spins, with a set of two spins between every set of three spins. In addition, this configuration is symmetric under a rotation of $90\dg$ about the conic axis. It appears that as one approaches the phase boundary from the ferromagnetic side the four sets of three spins become the 12 spins which point in $\langle110\rangle$ directions in the Cuboctahedral Stack state, and the four sets of two spins become the eight spins which point in $\langle111\rangle$ directions in the Cuboctahedral Stack state. 

This evidence suggests that the phase transition between the ferromagnetic and Cuboctahedral Stack states is continuous (of second order) and that the spin configuration near the phase boundary (on the ferromagnetic side) is a configuration which smoothly interpolates between the ferromagnetic and Cuboctahedral Stack states. 

\subsection{Cuboctahedral Stack in the $J_1$, $J_3$, $J_3'$ Phase Diagram}
\label{sec:CSJ3}

The Cuboctahedral Stack state also exists in the $J_1$, $J_3$, $J_3'$ phase diagram for $J_1$, $J_3 > 0$ and $J_3' < 0$. This state is again made up of three of the four $\left\{\tfrac{1}{2}\tfrac{1}{2}\tfrac{1}{2}\right\}$ type wavevectors, with the fourth unused wavevector pointing in the stacking direction. The parameterization is exactly the same as that presented in equations \eqref{CS0} and \eqref{CS123}. The energy per site for the Cuboctahedral Stack state with these three interactions is
	\beq
		\Eavg = -\left(\frac{3}{4} + \frac{\sqrt{6}}{2}\right)J_1 - J_3 + J_3'\ .
	\eeq

This state is expected to transition to the ferromagnetic state, with energy per site \eqref{EferroJ3}, when
	\beq
		J_1 = -\frac{8J_3 + 16J_3'}{9-2\sqrt{6}}\ .
	\eeq

From our experience with the phase transition between the ferromagnetic and Cuboctahedral Stack states in the $J_1$-$J_2$ phase diagram, we expect that this phase transition will also be of second order.

Previously, we had understood the Cuboctahedral Stack as being the result of competition between a strong ferromagnetic nearest neighbor interaction $J_1$ and a weak antiferromagnetic second nearest neighbor interaction $J_2$. In this case we have no second nearest neighbor interaction, but we do have a ferromagnetic $J_3$ interaction and an antiferromagnetic $J_3'$ interaction. In the Cuboctahedral Stack configuration, a spin in a triangular lattice layer has a dot product of $\tfrac{1}{3}$ with its six $J_3$ neighbors and a dot product of $-\tfrac{1}{3}$ with its six $J_3'$ neighbors. A spin in a Kagome lattice layer has a dot product of 1 with two of its six $J_3$ neighbors (those in adjacent Kagome lattice layers) and a dot product of 0 with the other four (those in the same Kagome lattice layer) and it also has a dot product of $-1$ with two of its $J_3'$ neighbors (those in the same Kagome lattice layer) and 0 with the other four (those in adjacent Kagome lattice layers). In this way the Cuboctahedral Stack state serves as a compromise in which spins have positive (or zero) dot products with their $J_3$ neighbors and negative (or zero) dot products with their $J_3'$ neighbors. So the dot product between each pair of spins is either zero or has the sign that goes with the interaction between that pair of spins (e.g. dot products between $J_3$ neighbors are either zero or positive, which favors the ferromagnetic $J_3$ interaction).

\section{Kawamura States}
\label{KawamuraStates}

\SAVE{MFL looked at the optimal LT wavevector in Summer 2011: 
it actually drifts in along a diagonal to settle near (0.73,0.73,0) for J2 near abs(J1).}

In the region $J_1 < 0$, $0 < J_2 \leq 1.09\left|J_1\right|$ of the phase diagram,
we encounter a remarkable phase first identified in finite-temperature
simulations of Refs.~\onlinecite{Tsu07} and \onlinecite{Na07}, and further studied
in Refs.~\onlinecite{Che08} and \onlinecite{Oku11}.
This is actually a family of phases dominated by a certain kind of
$\{\frac{3}{4}\frac{3}{4}0\}$ ordering mode,
which according to the Luttinger-Tisza analysis
is optimal  in this parameter range. In fact, the exact Luttinger-Tisza wavevector starts out at
$\{\frac{3}{4}\frac{3}{4}0\}$ for $J_2 \ll |J_1|$, but it drifts slightly inwards as $J_2$ is increased,
eventually settling at an incommensurate wavevector near $2\pi(0.73,0.73,0)$ as $J_2$ is increased 
to $J_2 \approx |J_1|$. So the $\{\frac{3}{4}\frac{3}{4}0\}$ wavevectors are only \emph{nearly} optimal 
in this region of parameter space, although they are very close to the true L.T. wavevector.
By taking linear combinations with multiple symmetry-related modes 
it is possible to construct three 
particularly symmetric noncoplanar states~\cite{Oku11}
that are called the ``Sextuplet-q'' state, and the ``Quadruplet-q states of types 1 and 2''.
We have called this whole family the ``Kawamura States" after the group 
which has done the most to elucidate their structure.

%

The Sextuplet-q state is composed from equal amounts of all six $\left\{\tfrac{3}{4}\tfrac{3}{4}0\right\}$ wavevectors. 
The Quadruplet-q state of type 1 is composed from {\it equal} amounts of 
four of the $\left\{\tfrac{3}{4}\tfrac{3}{4}0\right\}$ wavevectors 
which lie on the points of a cross in reciprocal space. 
[For example, these might be $2\pi(\tfrac{3}{4},\tfrac{3}{4},0)$, $2\pi(\tfrac{3}{4},-\tfrac{3}{4},0)$, $2\pi(\tfrac{3}{4},0,\tfrac{3}{4})$ and $2\pi(\tfrac{3}{4},0,-\tfrac{3}{4})$.]
Finally, the Quadruplet-q state of type 2 is composed of {\it unequal}
amounts of those same four wavevectors. [For example, the largest
contribution might be from wavevectors 
$2\pi(\tfrac{3}{4},\tfrac{3}{4},0)$ and $2\pi(\tfrac{3}{4},-\tfrac{3}{4},0)$,
with a slightly smaller contribution from $2\pi(\tfrac{3}{4},0,\tfrac{3}{4})$ and $2\pi(\tfrac{3}{4},0,-\tfrac{3}{4})$.]
Furthermore, each of these three states very nearly satisfies
the tetrahedron constraint ~\eqref{eq:tetzero}, so 
these states are a subset of the pure $J_1 < 0$ antiferromagnet 
of section \ref{J1}.

Ref.~\onlinecite{Oku11} used mean-field theory to predict the stability
of these phases as a function of temperature, concentrating on
the particular point $J_2/|J_1|=0.2$, and found that the cubic Sextuplet-q
phase is stable at higher temperatures, while the quadruplet-q kind
of state was stable at lower temperatures.~\footnote{
Ref.~\onlinecite{Che08} found that at even higher temperatures,
a collinear state based on the same modes is stabilized.}
In our $T=0$ study, the 
three states were nearly degenerate in this region of parameter space. 
The L.T. phase diagram indicates that the Kawamura states should transition to the
ferromagnetic state at $J_2 \approx 1.09\left|J_1\right|$. Iterative Minimization
simulations confirm that a phase transition does take place at this location and the 
simulations also seem to show that the transition is of second-order.
\SAVE{As determined from LT phase diagram and from IM simulations. 
But MFL's simulation data seems to point towards a
second order transition.}

Our Iterative Minimization simulations in this region of the $J_1$-$J_2$ phase diagram find these three states, as well as states with slightly higher energies which are also mainly composed of $\left\{\tfrac{3}{4}\tfrac{3}{4}0\right\}$ wavevectors. Therefore we suspect that the energy landscape in this region of parameter space consists of three local minima representing the three Kawamura states, and that these minima sit in a valley with a gently sloping floor. This would explain why our simulations sometimes don't find these three local minima and instead settle into states with slightly higher energies.

Even though the three Kawamura states are nearly degenerate in this region of the phase diagram, our Iterative Minimization simulations show that the true ground state is the Sextuplet-q state for $J_2 \ll |J_1|$ at least until $J_2 = 0.1|J_1|$. Somewhere between $J_2 = 0.1|J_1$ and $J_2 = 0.15|J_1|$ there is a phase transition, and we find that the true ground state for $0.15|J_1| \leq J_2 \leq 1.09|J_1|$ is the Quadruplet-q state of type 2. It appears that the Quadruplet-q state of type 1 is never the true ground state in this region, even though it is always nearly degenerate with the other two Kawamura states.  



We have focused our study of the structure of the states in this region almost exclusively on the Kawamura Sextuplet-q state. There are two reasons for this. Firstly, it has the most symmetry in real space and in spin space 
out of the three Kawamura states. More importantly, it has a slightly 
lower energy than the other two Kawamura States when 
$J_2 \ll \left|J_1\right|$, suggesting that it is the state selected 
out of the highly degenerate manifold of pure-$J_1$ ground
states by an infinitesimal ferromagnetic $J_2$ interaction.

\subsection{Structure of Kawamura Sextuplet-q State}
\label{KawamuraSq}

The Kawamura Sextuplet-q state is a complicated non-coplanar state which is mostly composed of equal amounts of the six $\left\{\tfrac{3}{4}\tfrac{3}{4}0\right\}$ wavevectors. Because of the complexity of this state, in this section we only present the results of analysis on a numerical spin configuration generated by an Iterative Minimization simulation with the parameter values $J_1 = -1$, $J_2 = 0.1$ on a lattice of size $4\times 4\times 4$ with periodic boundary conditions. 

In this state, it turns out that the site energy $\Esite$ of the spins takes on only six different values: thus we can organize the spins by site energy into six sublattices in real space,
each of which has the periodicity of a $2\times2\times2$ 
supercell (and is not itself a Bravais lattice).
In fact, the sites in each sublattice are equivalent by symmetries of the
ground state.
The site energy and number of spins in each sublattice is given 
in table \ref{KawaTab1}; the site positions for each sublattice 
are given in Appendix \ref{KawaSub}.

	\begin{table}[t]
	\begin{center}
	  \begin{tabular}{ | c | c | c |}
	\hline
	 Sublattice & $\Esite$   & \# of spins \\ \hline
	 1 & -1.1053 & 192 \\ \hline
	 2 & -1.1841 & 192 \\ \hline
	 3 & -1.2103 & 192 \\ \hline
	 4 & -1.2672 & 64 \\ \hline
	 5 & -1.2753 & 192 \\ \hline
	 6 & -1.2902 & 192 \\ 
	    \hline
	  \end{tabular}
	\caption{Site energies and number of spins in each sublattice of the Kawamura Sextuplet-q State with $J_1 = -1$, $J_2 = 0.1$, on a lattice of size $4\times 4\times 4$.} 
	\label{KawaTab1}
	\end{center}
	\end{table}

Sublattices 3 and 4 are the sublattices of highest symmetry in this configuration. They are each made up of tetrahedra of spins. The tetrahedra in sublattice 4 sit on the sites of a body-centered cubic (B.C.C.) lattice with a unit cell of size $2\times2\times2$. The tetrahedra of spins in sublattice 3 combine with those in sublattice 4 to create a simple cubic lattice of tetrahedra with a unit cell of size $1\times1\times1$. The spins in sublattice 3 point towards the 12 edge-midpoints of a cube in spin space and the spins in sublattice 4 point towards the eight corners of that same cube. 

The nearest neighbors of the spins in sublattices 3 and 4 are distributed throughout the six sublattices. Because sublattices 3 and 4 are made up of tetrahedra, three of the six nearest neighbors of each spin in these two sublattices lie in the same sublattice as that spin. The other three nearest neighbors of the spins in sublattice 4 (there are 192 = 3$\times$64 of them) are contained in sublattice 5 (these are the only spins in sublattice 5). The other three nearest neighbors of each of the spins in sublattice 3 (there are 576 = 3$\times$192 of them) are distributed throughout sublattices 1, 2 and 6.

Next we present the results of a least-squares fit of this numerical spin configuration to functions of the form $c_j\sin(\mb{q}_j\cdot\mb{r} + \phi_j)$ where $\mb{q}_j$ is one of the wavevectors of type $\left\{\tfrac{3}{4}\tfrac{3}{4}0\right\}$ (an optimal L.T. wavevector) and the $c_j$ and $\phi_j$ are the undetermined fitting parameters. We use the condensed notation
	\begin{subequations}
	\label{condensed}
	\beqa
	   \Phi_{xy} &=& 2\pi\left(\frac{3}{4}x + \frac{3}{4}y\right) \\
	   \bar{\Phi}_{xy} &=& 2\pi\left(\frac{3}{4}x - \frac{3}{4}y\right),
	\eeqa
	\end{subequations}
and a similar notation for phase functions involving $x$ and $z$ or 
involving $y$ and $z$, to simplify the presentation of this state. 
Idealizing the results of the least square fit gives an approximate parameterization of this state in each F.C.C. sublattice of the form 
	\beqa
		\mb{S}_{\al} \ = \ \left\{ A_{\al}\sin\left(\Phi_{yz} - \frac{\pi}{8}\right) + B_{\al}\sin\left(\bar{\Phi}_{yz}\right) \right\}\xhat \nonumber \\ 
			+ \ \left\{ C_{\al}\sin\left(\Phi_{xz} + \frac{\pi}{8}\right) + D_{\al}\sin\left(\bar{\Phi}_{xz} + \frac{\pi}{4}\right) \right\}\yhat \nonumber \\
			+ \ \left\{ E_{\al}\sin\left(\Phi_{xy} + \frac{\pi}{8}\right) + F_{\al}\sin\left(\bar{\Phi}_{xy} + \frac{\pi}{4}\right) \right\}\zhat \label{Kawa1}
	\eeqa
where the values of the coefficients $A_{\al}$, $B_{\al}$, $C_{\al}$, $D_{\al}$, $E_{\al}$ and $F_{\al}$ are given for each sublattice in table \ref{KawaCoeffs}.

\LATER{12
About \eqref{Kawa1}: CLH has not fully digested the symmetry structure,
but clearly  the $yz$, $xz$, and $xy$ kinds of term can be
associated with $x$, $y$, and $z$ directions respectively.
So I wonder about possibly changing the notation to highlight this.
For example, in place of $A, C, E$ maybe $A, B, C$; in place
of $B, D, F$, maybe $\bar{A}, \bar{B}, \bar{C}$?}

We can see from \eqref{Kawa1} that if one fixes the x and y coordinates and moves in the z-direction, then the spin configuration looks roughly like a superposition of two distorted conic spirals about the z-axis. A similar statement holds if one fixes the x and z coordinates and moves in the y-direction or fixes the y and z coordinates and moves in the x-direction.

The energy per site for this state is $\Eavg = -1.2164$.

\begin{table}[t]
	\begin{center}
	  \begin{tabular}{ | c | c | c | c | c | c | c |}
	\hline
	 $\al$ & $A_{\al}$ & $B_{\al}$  & $C_{\al}$ & $D_{\al}$ & $E_{\al}$ & $F_{\al}$ \\ \hline
	 1 &$-c_1$ & $c_2$ & $-c_1$ & $-c_2$ & $c_1$ & $-c_2$ \\ \hline
	 2 & $-c_1$ & $c_2$ & $c_2$ & $c_1$ & $-c_2$ & $c_1$  \\ \hline
	 3 & $c_2$ & $-c_1$ & $-c_1$ & $-c_2$ & $-c_2$ & $c_1$ \\ \hline
	 4 & $c_2$ & $-c_1$ & $c_2$ & $c_1$ & $c_1$ & $-c_2$ \\ 
	    \hline
	  \end{tabular}
	\caption{Coefficients for the approximate parameterization \eqref{Kawa1} of the Kawamura Sextuplet-q state. The greek letter $\al$ indexes the four F.C.C. sublattices and the coefficients are all given in terms of the two constants $c_1 \approx 0.73$ and $c_2 \approx 0.27$.
\SAVE{Is is possible $c_2=1-c_1$, or not quite? MFL is not really sure, 
but doesn't think so since these coefficients are averages of a couple 
different but close values.}}

	\label{KawaCoeffs}
	\end{center}
	\end{table}

\subsection{Anharmonic selection as a consequence of normalization}

Although the parameterization \eqref{Kawa1} looks relatively clean, it is not normalized on all of the lattice sites. This is because the Kawamura Sextuplet-q state is not constructed solely from $\left\{\tfrac{3}{4}\tfrac{3}{4}0\right\}$ wavevectors. A closer look at the Fourier Transform of the numerical spin configuration found in Iterative Minimization simulations shows that it also contains small contributions from wavevectors like $2\pi(\tfrac{1}{4},\tfrac{1}{2},\tfrac{3}{4})$, $2\pi(\tfrac{1}{4},\tfrac{1}{2},-\tfrac{3}{4})$ and $2\pi(\tfrac{1}{2},\tfrac{1}{4},-\tfrac{1}{4})$, even though these are not optimal Luttinger-Tisza wavevectors. The first two of these wavevectors are equivalent by a symmetry of the F.C.C. lattice (reflection through the x-y plane), so they give the same eigenvalues for the matrix $\un{\JLT}(\mb{q})$. The third kind of wavevector is not equivalent to the first two.

It turns out that wavevectors of these three types are actually just linear combinations of three of the optimal L.T. wavevectors. For  example, we can write $2\pi(\tfrac{3}{4},-\tfrac{3}{4},0) + 2\pi(\tfrac{3}{4},-\tfrac{3}{4},0) + 2\pi(\tfrac{3}{4},0,\tfrac{3}{4}) = 2\pi(\tfrac{9}{4},-\tfrac{3}{2},\tfrac{3}{4})$. But we may add to this wavevector a reciprocal lattice vector of the form $2\pi(-2,2,0)$ to map it back to a wavevector in the first Brillouin zone. We see then that this linear combination of optimal L.T. wavevectors is equivalent to the wavevector $2\pi(\tfrac{1}{4},\tfrac{1}{2},\tfrac{3}{4})$. It turns out that the small contributions from these three kinds of wavevectors can be understood by looking at what happens to our approximate parameterization when we normalize it.

To normalize the parameterization \eqref{Kawa1}, we would compute
	\beq
		\what{\mb{s}}_i = \frac{\mb{s}_i}{\sqrt{\left|\mb{s}_i\right|^2}}
	\eeq
for every spin. We can rewrite the magnitude of the spin $\mb{s}_i$ as $\sqrt{\left|\mb{s}_i\right|^2} = \sqrt{1 + \left(\left|\mb{s}_i\right|^2 -1 \right)}$. Taylor expanding the reciprocal of this quantity about the point $\left|\mb{s}_i\right|^2 = 1$ (where the state is normalized), we find
	\beq
		\what{\mb{s}}_i = \mb{s}_i\left[ 1 -\frac{1}{2}\left(\left|\mb{s}_i\right|^2 -1 \right) + \dots\right]\ .
	\eeq
So the first order correction to the non-unit vector spin $\mb{s}_i$ includes a term $\left|\mb{s}_i\right|^2\mb{s}_i$. It is from this cubic term that we get corrections to the state that involve wavevectors which are a linear combination of three of the optimal Luttinger-Tisza wavevectors (if we wrote out $\mb{s}_i$ as an exponential Fourier series, we would see the addition of wavevectors happening in the exponents). 

\section{A New Variety of Double-Twist State}
\label{DoubleTwist}

In the region $J_1>0$, $-.68 \leq J_2 \leq -.43$, we find a state which is chiefly composed of a dominant set of two of the six
$\left\{\tfrac{3}{4}\tfrac{3}{4}0\right\}$ wavevectors and a sub-dominant set of two of the other wavevectors of this type, and
these four wavevectors all lie on the points of a cross in reciprocal space. For example, this state might be made from the wavevectors
$2\pi(\tfrac{3}{4},\tfrac{3}{4},0)$, $2\pi(\tfrac{3}{4},-\tfrac{3}{4},0)$, $2\pi(\tfrac{3}{4},0,\tfrac{3}{4})$ and $2\pi(\tfrac{3}
{4},0,-\tfrac{3}{4})$ with the Fourier amplitudes of the first two wavevectors being approximately $1.40$ times the Fourier amplitudes
of the second two wavevectors. Luttinger-Tisza analysis confirms that wavevectors of the type 
$\left\{\tfrac{3}{4}\tfrac{3}{4}0\right\}$ are the optimal wavevectors for this state. 

\SAVE{Interestingly, these modes have the same wavevectors 
as the modes which go into the Kawamura States discussed in 
section~\ref{KawamuraStates}, although these are not the same modes.}

Although this state is mostly constructed from $\left\{\tfrac{3}{4}\tfrac{3}{4}0\right\}$ wavevectors, it also contains small contributions from $\left\{\tfrac{3}{4}\tfrac{1}{4}0\right\}$ wavevectors. These extra wavevectors aid in normalizing this state, since one cannot construct a normalized state from $\left\{\tfrac{3}{4}\tfrac{3}{4}0\right\}$ wavevectors alone.

\LATER{13 MFL: Is this obvious? 
CLH: I'm sure no one else has derived this, and  I don't find it obvious
myself.  I ought to write up a general discussion of when one can
or can't normalize.  You can temporarily
leave it here as an ungrounded assertion.}

This state shares three characteristics in common with the 
Double-Twist state found on the Octahedral lattice in Ref.~\onlinecite{SH}:
\begin{itemize}
\item[(1)]
There is a distinguished direction $\widehat{\mb{m}}$ in real space and 
also a distinguished direction $\zhat$ in spin space. This state can be written using a basis for spin space that uses the fixed basis vector $\zhat$ and two rotating basis vectors $\widehat{\mb{A}}(\mb{r}\cdot\widehat{\mb{m}})$ and $\widehat{\mb{B}}(\mb{r}\cdot\widehat{\mb{m}})$ which lie in the plane perpendicular to $\zhat$, so that $\widehat{\mb{A}}(\mb{r}\cdot\widehat{\mb{m}})\times\widehat{\mb{B}}(\mb{r}\cdot\widehat{\mb{m}}) = \zhat$. As one moves in the $\widehat{\mb{m}}$ direction in real space the rotating basis vectors $\widehat{\mb{A}}(\mb{r}\cdot\widehat{\mb{m}})$ and $\widehat{\mb{B}}(\mb{r}\cdot\widehat{\mb{m}})$ spiral about the fixed vector $\zhat$ in spin space.
\item[(2)]
There is a second direction $\widehat{\mb{n}}$ in real space orthogonal to the $\widehat{\mb{m}}$ direction, with the property that the spins also spiral as one moves in the $\widehat{\mb{n}}$ direction, although not about the $\zhat$ axis in spin space. Since the rotating basis vectors stay put as one moves in the $\widehat{\mb{n}}$ direction, this second kind of spiral is completely independent from the first kind.
\item[(3)]
Both distinct kinds of spiraling behavior are controlled by the same type of wavevector.
\end{itemize}

To summarize, this state gets its name from the fact that the spins are tracing out two different kinds of spirals in two directions which are perpendicular to each other, but these two spirals are controlled by the same type of wavevector.
In the Double-Twist state that we find on the Pyrochlore lattice the distinguished direction $\what{\mb{m}}$ in real space is a $\langle100\rangle$ direction, which we take to be the z-direction in our discussion of this state. The second spiraling direction $\what{\mb{n}}$ is then the x- or y-direction. Both kinds of spiral are controlled by a $\left\{\tfrac{3}{4}00\right\}$ wavevector.

We now present an approximate parameterization of this state obtained by performing a least squares fit of a numerical spin configuration found in an Iterative Minimization simulation onto sines and cosines of $\mb{q}\cdot\mb{r}$, where $\mb{q}$ is a $\left\{\tfrac{3}{4}\tfrac{3}{4}0\right\}$ wavevector (an optimal L.T. wavevector). This simulation was performed with the parameter values $J_1 = 1$ and $J_2 = -0.6$ on a lattice of size $4\times 4\times 4$ using periodic boundary conditions. 

The two dominant $\left\{\tfrac{3}{4}\tfrac{3}{4}0\right\}$ wavevectors making up this state are $2\pi(\tfrac{3}{4},\tfrac{3}{4},0)$ and $2\pi(\tfrac{3}{4},-\tfrac{3}{4},0)$ and the two sub-dominant wavevectors making up this state are $2\pi(\tfrac{3}{4},0,\tfrac{3}{4})$ and $2\pi(\tfrac{3}{4},0,-\tfrac{3}{4})$. 
As in \eqref{condensed} we use the condensed notation
	\begin{subequations}
	\label{condensedDT}
	\beqa
		\Phi_{xy} &=& 2\pi\left(\frac{3}{4}x + \frac{3}{4}y\right) \\
		\bar{\Phi}_{xy} &=& 2\pi\left(\frac{3}{4}x - \frac{3}{4}y\right) \\
		\Phi_x &=& 2\pi\left(\frac{3}{4}x\right)
	\eeqa
	\end{subequations}
with similar notations for arguments involving y and z. We express this state in terms of the two rotating basis vectors
	\begin{subequations}
	\beqa
		\what{\mb{A}}(z) &=& \cos\left(\Phi_z\right)\xhat + \sin\left(\Phi_z\right)\yhat \\
		\what{\mb{B}}(z) &=& -\sin\left(\Phi_z\right)\xhat + \cos\left(\Phi_z\right)\yhat \ ,
	\eeqa
	\end{subequations}
which have been chosen so that $\what{\mb{A}}(z)\times\what{\mb{B}}(z) = \zhat$. The approximate parameterization of this state in each F.C.C. sublattice has the form
	\beqa
		\mb{S}_{\al} \ = \ \left\{ A'_{\al}\sin\left(\Phi_y + \theta_{\al}\right) + B'_{\al}\sin\left(\Phi_y + \psi_{\al}\right) \right\}\what{\mb{A}}(z) \nonumber \\
+ \ \left\{ C'_{\al}\cos\left(\Phi_y + \theta_{\al}\right) + D'_{\al}\cos\left(\Phi_y + \psi_{\al}\right) \right\}\what{\mb{B}}(z) \nonumber \\
+ \ \left\{ E'_{\al}\sin\left(\Phi_{xy} + \frac{\pi}{4}\right) + F'_{\al}\sin\left(\bar{\Phi}_{xy} + \frac{\pi}{8}\right) \right\}\zhat \label{DT1}
	\eeqa
where the values of the coefficients $A'_{\al}$, $B'_{\al}$, $C'_{\al}$, $D'_{\al}$, $E'_{\al}$ and $F'_{\al}$ and the phase angles $\theta_{\al}$ and $\psi_{\al}$ for each sublattice are given in table \ref{DTCoeffs}.


\begin{table}[t]
	\begin{center}
	  \begin{tabular}{ | c | c | c | c | c | c | c | c | c |}
	\hline
	 $\al$ & $A'_{\al}$ & $B'_{\al}$  & $C'_{\al}$ & $D'_{\al}$ & $E_{\al}$ & $F'_{\al}$ & $\theta_{\al}$ & $\psi_{\al}$  \\ \hline
	 1 & $d_1$ & $d_2$ & $-d_1$ & $d_2$ & $d_3$ & $d_4$ & 
                            $5.0\dg$ & $3.8\dg$ \\ \hline
	 2 & $d_1$ & $d_2$ & $-d_1$ & $d_2$ & $d_4$ & $d_3$ & 
                            $31.0\dg$ & $4.5\dg$ \\ \hline
	 3 & $d_2$ & $d_1$ & $-d_2$ & $d_1$ & $d_4$ & $d_3$ & 
                            $17.7\dg$ & $-7.4\dg$ \\ \hline
	 4 & $d_2$ & $d_1$ & $-d_2$ & $d_1$ & $d_3$ & $d_4$ & 
                            $18.4\dg$ & $17.3\dg$ \\ \hline
	  \end{tabular}
	\caption{Coefficients and phase angles for the approximate parameterization \eqref{DT1} of the Double-Twist state. The greek letter $\al$ indexes the four F.C.C. sublattices and the coefficients are all given in terms of the four constants $d_1 \approx 0.23$, $d_2 \approx 0.76$, $d_3 \approx 0.26$, $d_4 \approx 0.75$. Analysis of multiple examples of this state seems to indicate that $d_1\neq d_3$ and $d_2\neq d_4$, even though their values appear to be very similar.}

	\label{DTCoeffs}
	\end{center}
	\end{table}

We see from \eqref{DT1} that if one holds $z$ constant and moves in the y-direction, the spins spiral (but not about the $\zhat$ axis, since the $\zhat$ component of the spins changes as one moves in the y-direction), and if one holds $y$ constant and moves in the z-direction the spins spiral about the $\zhat$ direction (since the basis vectors $\what{\mb{A}}(z)$ and $\what{\mb{B}}(z)$ change as one moves in the z-direction). In addition, these two spirals are controlled by a $\left\{\tfrac{3}{4}00\right\}$ wavevector. So in this example the $\what{\mb{m}}$ direction is the z-direction and the $\what{\mb{n}}$ direction is the y-direction. 

This state has an energy per site $\Eavg = -2.2780$. 

\section{Multiply-Modulated Commensurate Spirals}
\label{MMCS}

	In the small sliver of parameter space $J_1 = 1$, $-0.43 \leq J_2 \leq -0.40$ we find a non-coplanar state with period 4 constructed solely from wavevectors of the types $\left\{\tfrac{3}{4}\tfrac{1}{4}\tfrac{1}{2}\right\}$, $\left\{\tfrac{3}{4}\tfrac{3}{4}0\right\}$ and $\left\{\tfrac{1}{4}\tfrac{1}{4}0\right\}$ (recall that these three types of wavevector are not equivalent by any symmetry of the F.C.C. lattice). Our prototype of this state uses the four wavevectors $\mb{q}_{1+} = 2\pi(\tfrac{3}{4},\tfrac{1}{4},\tfrac{1}{2})$, $\mb{q}_{1-} = 2\pi(\tfrac{3}{4},\tfrac{1}{4},-\tfrac{1}{2})$, $\mb{q}_2 = 2\pi(\tfrac{3}{4},-\tfrac{3}{4},0)$, and $\mb{q}_3 = 2\pi(\tfrac{1}{4},-\tfrac{1}{4},0)$, but one can build this state from any set of four wavevectors that is equivalent to these by symmetry. The $\left\{\tfrac{3}{4}\tfrac{1}{4}\tfrac{1}{2}\right\}$ wavevectors are the optimal Luttinger-Tisza wavevectors in this region of parameter space. In fact, the largest eigenvalues of the matrix $\mb{\JLT}(\mb{q})$ for these wavevectors at the parameter values $J_1 = 1$, $J_2 = -0.42$ are $\lam_{max}(\mb{q}_{1\pm}) = \LTeig = 2.0195$, $\lam_{max}(\mb{q}_2) = 2.0121$ and $\lam_{max}(\mb{q}_3) = 0.9337$. The Luttinger-Tisza wavevectors $\mb{q}_{1\pm}$ are the dominant wavevectors in this state, but the state also contains sizable contributions from modes containing the sub-optimal wavevectors $\mb{q}_2$ and $\mb{q}_3$ (see equations \eqref{MMCS03} and \eqref{MMCS12}). So this state is another example of a non-coplanar state which is constructed using an optimal L.T. wavevector and a set of wavevectors which are suboptimal.

\begin{figure}[t]
  \caption{Largest eigenvalue $\lam_{max}(\mb{q})$ of the matrix $\un{\JLT}(\mb{q})$ vs. $J_2/J_1$ for the optimal L.T. wavevector of each of the five ground states found in the region $J_1 > 0$, $J_2 < 0$. The name of the corresponding ground state is shown in parentheses next to the label for the type of wavevector. The five ground states are the Ferromagnetic (F.M.), Cuboctahedral Stack (C.S.), Multiply-Modulated Commensurate Spiral (M.M.C.S.), Double-Twist (D.T.) and $\left\{q00\right\}$ Planar Spiral states (with wavenumber $q$ for the Planar Spiral given by equation \eqref{Qspiral}). The inset shows the small region $-0.43 \leq J_2/J_1 \leq -0.40$ where the largest eigenvalue for the dominant wavevector in the Multiply-Modulated Commensurate Spiral state briefly crosses above the largest eigenvalue for the wavevectors characterizing the Cuboctahedral Stack and Double-Twist states, confirming that there is a distinct ground state in this small region of the $J_1$-$J_2$ phase diagram.}
\vskip 10pt
  \centering
    \includegraphics[width= .5\textwidth]{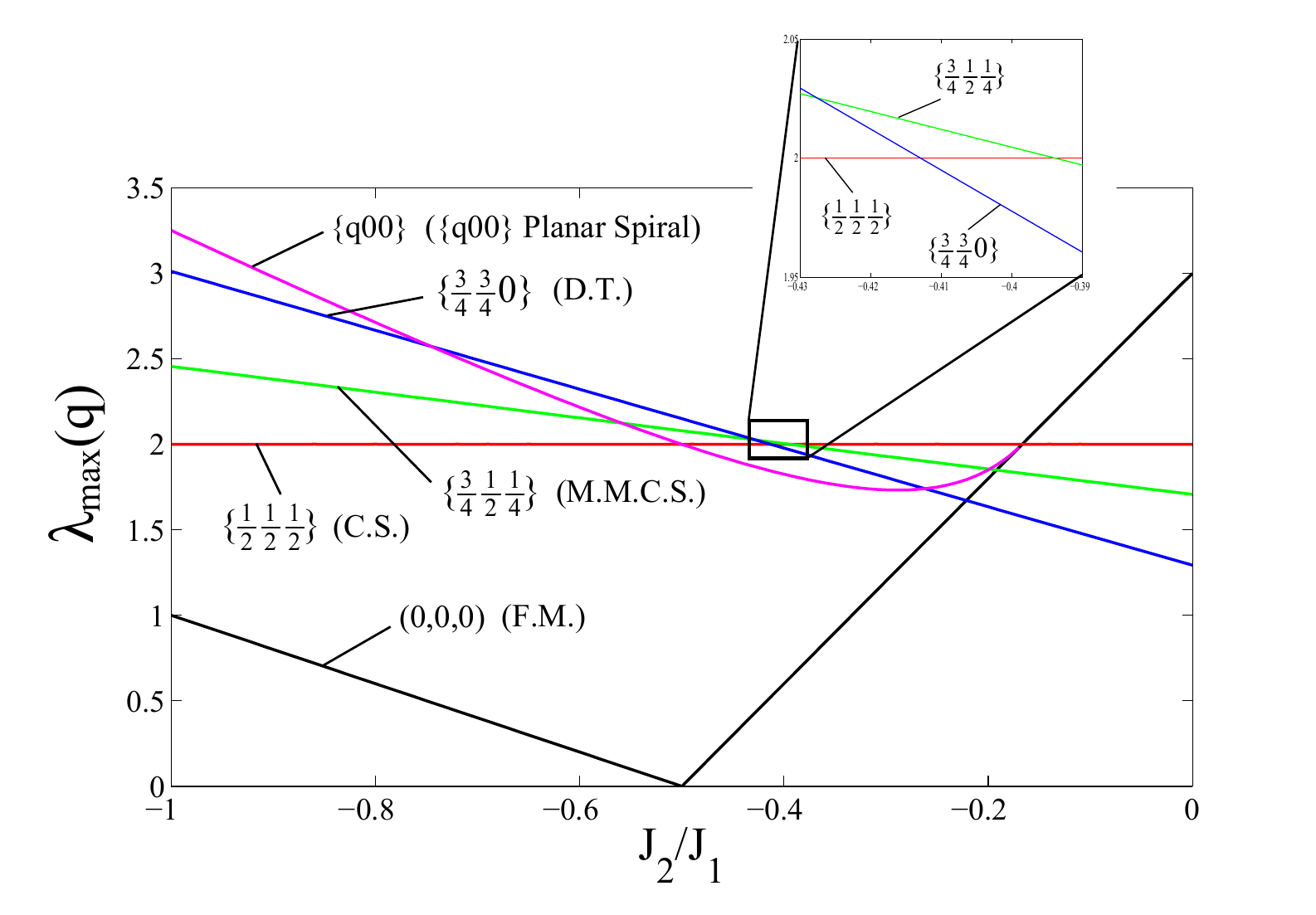}
\label{fig:J1posJ2neg}
\end{figure}

Figure \ref{fig:J1posJ2neg} shows the largest eigenvalue $\lam_{max}(\mb{q})$ of the matrix $\un{\JLT}(\mb{q})$ for the optimal Luttinger-Tisza
wavevector of each of the five states found in the region $J_1>0$, $J_2<0$ as a function of $J_2/J_1$. It also shows a zoomed in view of
the region $-0.43 \leq J_2/J_1 \leq -0.40$ where the largest eigenvalue for wavevectors of the type
$\left\{\tfrac{3}{4}\tfrac{1}{4}\tfrac{1}{2}\right\}$, the main type of wavevector in the Multiply-Modulated Commensurate Spiral state,
 briefly crosses above the largest eigenvalue of $\un{\JLT}(\mb{q})$ for the wavevectors used in the Cuboctahedral Stack and
Double-Twist states. 
The fact that the L.T. phase diagram shows a sliver in which $\left\{\tfrac{3}{4}\tfrac{1}{4}\tfrac{1}{2}\right\}$ wavevectors are
optimal is strong evidence that the corresponding spin state found in
Iterative Minimization simulations is not merely an artifact of relaxation.
\LATER{13a My guess would be the other two wavevectors are possible because they are NEARLY
degenerate; it was speculated in (Sophia's paper) that these admixtures tend to arise 
not far from the parameters where two LT modes become accidentally degenerate.
It looks that way in the figure, but one of the lines is CS,  i.e. (1/2 1/2 1/2),
which isn't one of the admixtures.  
The other is labeled DT, made from (3/4 3/4 0) which is indeed one of the admixed
wavevectors.}

We call this state a Multiply-Modulated Commensurate Spiral (M.M.C.S.) because the spins in this state are tracing out two different spirals in two orthogonal directions \emph{and} these two spirals are controlled by two different kinds of wavevectors. 
\SAVE{(The presence of two different kinds of wavevectors which control the spirals is what gives this state the descriptor ``Multiply-Modulated". It is a commensurate spiral because every wavevector involved is commensurate with the lattice.)}

\LATER{(This comment from 10/23 -- CLH needs to check if it was
implemented.)
My main change in this section is that the presentation $\mb{r}\cdot\hat{\mb{m}}$
is opaque.  You don't seem to come out and say this special direction is just
one of the $\mb{q}$ vectors.  I think it would be better to replace the argument
by a single coordinate, $\hat{\mb{A}}(\mb{r}\cdot\hat{\mb{m}}) \to 
\hat{\mb{A}}(\xi)$ and say -- once -- that ``$\xi\equiv \mb{r}\cdot \mb{q}_3$''.
Note also I'm not thrilled with using the unit vector of $\mb{q}_3$ to define
the coordinate, which just introduces confusing factors of $\sqrt{10}$; if we use
the wavevector itself, the coordinate values will be integers or at least
multiples of 1/4.  Right?}

In this state there is a distinguished direction $\widehat{\mb{m}}$ in real space and a distinguished direction $\zhat$ in spin space. 
The distinguished direction in real space is parallel to $\mb{q}_{1+} + \mb{q}_{1-}$, so $\widehat{\mb{m}}$ points in the $[310]$
direction in our example state. 
The spin configuration can be written using a basis in spin space that uses the one fixed basis vector $\zhat$ and two rotating basis vectors $\widehat{\mb{A}}(\xi)$ and $\widehat{\mb{B}}(\xi)$, where the coordinate $\xi = \mb{r}\cdot\widehat{\mb{m}}$. These two basis vectors lie in the plane perpendicular to $\zhat$ and satisfy $\widehat{\mb{A}}(\xi)\times\widehat{\mb{B}}(\xi) = \zhat$. As one moves in the $\widehat{\mb{m}}$ direction in real space these two basis vectors rotate counter-clockwise about $\zhat$ in spin space. To write this state in the most concise way we also use a second set of two rotating basis vectors $\widehat{\mb{A}}'(\xi)$ and $\widehat{\mb{B}}'(\xi)$ lying in the plane perpendicular to $\zhat$ and also satisfying
$\widehat{\mb{A}}'(\xi)\times\widehat{\mb{B}}'(\xi) = \zhat$, but this second set rotates
clockwise about $\zhat$ as one moves in the $\widehat{\mb{m}}$ direction in real space.

There is also a second distinguished direction $\widehat{\mb{n}}$ in real space which is parallel to $\mb{q}_{1+} - \mb{q}_{1-}$,so $\widehat{\mb{n}}$ points in the $[001]$ direction in our example state. Because $\mb{q}_{1+}$ and $\mb{q}_{1-}$ differ only in the sign of one component, we have $\widehat{\mb{m}}\cdot\widehat{\mb{n}} = 0$. In this state the spins also spiral about about the $\zhat$ direction in spin space when one moves in the $\what{\mb{n}}$ direction in real space, even though the four basis vectors $\widehat{\mb{A}}(\xi)$, $\widehat{\mb{B}}(\xi)$, $\widehat{\mb{A}}'(\xi)$ and $\widehat{\mb{B}}'(\xi)$ stay put as one moves in this direction in real space. Finally, the wavevectors that control the spiraling in the $\widehat{\mb{m}}$ and $\widehat{\mb{n}}$ directions are not equivalent by symmetry, so the spiraling behavior in this state is different from the spiraling behavior of the Double-Twist state of section \ref{DoubleTwist} (the two kinds of spirals in the Double-Twist state are controlled by the same type of wavevector). 

Using a diagnostic which counts the number of different spin directions present in a certain state (two spins $\mb{s}_i$ and $\mb{s}_j$ are considered to point in different directions if $\mb{s}_i\cdot\mb{s}_j < .99$), we find that the spins in this state point in 42 different directions. The spins in two of the F.C.C. sublattices point in the same 10 directions and the spins in the other two F.C.C. sublattices point in 32 directions which are distinct from the first 10.

Here we present an approximate parameterization of this state which was constructed by idealizing the result of a least square fit of a numerical spin configuration to sines and cosines of $\mb{q}\cdot\mb{r}$, where $\mb{q}$ is a $\left\{\tfrac{3}{4}\tfrac{1}{4}\tfrac{1}{2}\right\}$, $\left\{\tfrac{3}{4}\tfrac{3}{4}0\right\}$, or a $\left\{\tfrac{1}{4}\tfrac{1}{4}0\right\}$ wavevector. This numerical spin configuration was generated by an Iterative Minimization simulation with the parameter values $J_1 = 1$ and $J_2 = -0.43$ on a lattice of size $4\times 4\times 4$ using periodic boundary conditions. 

The approximate parameterization uses the four wavevectors $\mb{q}_{1+}$, $\mb{q}_{1-}$, $\mb{q}_2$, and $\mb{q}_3$ defined in the first paragraph of this section. We again use the condensed notation 
	\beqa
		\Phi_{xy} &=& 2\pi\left(\tfrac{3}{4}x + \tfrac{1}{4}y - \tfrac{1}{16}\right) \\
		\Theta_{xy} &=& 2\pi\left(\tfrac{3}{4}x - \tfrac{3}{4}y - \tfrac{1}{16}\right) \\
		\Psi_{xy} &=& 2\pi\left(\tfrac{1}{4}x - \tfrac{1}{4}y - \tfrac{3}{16}\right)\ .
	\eeqa
as a shorthand for these position dependent phase angles. The most concise presentation of this state uses two sets of two rotating basis vectors, all lying in the plane in spin space perpendicular to $\zhat$. We can write these four basis vectors as
	\begin{subequations}
	\beqa
		\what{\mb{A}}(x,y) &=& \cos \Phi_{xy} \xhat + \sin \Phi_{xy} \yhat \label{Axy}\\
		\what{\mb{B}}(x,y) &=& -\sin \Phi_{xy} \xhat + \cos \Phi_{xy} \yhat \\
		\what{\mb{A}}'(x,y) &=& \cos \Phi_{xy} \xhat - \sin \Phi_{xy} \yhat \\
		\what{\mb{B}}'(x,y) &=& \sin \Phi_{xy} \xhat + \cos \Phi_{xy} \yhat \label{Bpxy}
	\eeqa	
	\end{subequations}
where we have arranged it so that $\what{\mb{A}}(x,y)\times\what{\mb{B}}(x,y) = \what{\mb{A}}'(x,y)\times\what{\mb{B}}'(x,y) = \ahat\times\bhat  = \zhat$. $\what{\mb{A}}'(x,y)$ is the reflection of $\what{\mb{A}}(x,y)$ over the $\xhat$ axis and $\what{\mb{B}}'(x,y)$ is the reflection of $\what{\mb{B}}(x,y)$ over the $\yhat$ axis. This parameterization takes quite different forms in the different F.C.C. sublattices. In sublattices 0 and 3, we have
	\beqa
		\mb{S}_{\al = 0,3} &=& l_1\left\{\sin(\pi z)\what{\mb{A}}'(x,y) \mp \cos(\pi z)\what{\mb{B}}'(x,y)\right\} \nonumber \\
		&+& l_2\left\{ \pm\sin(\pi z)\what{\mb{A}}(x,y) - \cos(\pi z)\what{\mb{B}}(x,y)\right\} \nonumber \\
		&+& \left\{ l_3\sin\Theta_{xy} + l_4\sin\Psi_{xy} \right\}\zhat \ , \label{MMCS03}
	\eeqa
where the top (bottom) sign in the plus/minus and minus/plus signs corresponds to the zeroth (third) F.C.C. sublattice and the coefficients are $l_1 \approx 0.868$, $l_2 \approx 0.054$, $l_3 \approx 0.699$ and $l_4 \approx 0.076$. These are the two sublattices in which the spins point in 32 different directions, as discussed above. In sublattices 1 and 2, the spin configuration has the same form, 
	\beqa
		\mb{S}_{\al = 1,2} &=& 2l_3\cos(\pi z)\what{\mb{B}}'(x,y)  \nonumber \\
		&+& \left\{ l_5\sin\Theta_{xy} + l_6\sin\Psi_{xy} \right\}\zhat \ , \label{MMCS12}
	\eeqa
where $l_5 \approx 0.228$, $l_6 \approx 0.011$ and $l_3$ is the same coefficient that appeared in \eqref{MMCS03}. These are the two sublattices in which the spins point in the same 10 directions. 


In this parameterization $\widehat{\mb{m}}$ points in the $[310]$ direction and $\widehat{\mb{n}}$ points in the $[001]$ direction.
Furthermore, the spiraling in the $\widehat{\mb{m}}$ direction is controlled by the wavevector $2\pi(\tfrac{3}{4},\tfrac{1}{4},0)$ and
the spiraling in the $\widehat{\mb{n}}$ direction is controlled by the wavevector $2\pi(0,0,\tfrac{1}{2})$, and these two wavevectors
are clearly not of the same type.

The energy per site for this state is $\Eavg = -2.0131$.

\section{Conclusion}

In conclusion, we have surveyed the full range of the phase diagram
of classical ground states with nearest and second-nearest neighbor
exchange couplings in the Pyrochlore lattice 
(Figure~\ref{fig:PhaseDiag} and Table~\ref{J1J2States}).
Large swathes of the phase diagram (roughly those in which 
$|J_2| > |J_1|$) are dominated by the ferromagnet or a 
relatively simple coplanar spiral phase, but with smaller
$|J_2|$ values (for either sign of $J_1$), a zoo of 
complex non-coplanar phases were found.

In a similar study of the Octahedral lattice, which formed the
model for this one, three categories of non-coplanar state
were identified,
with only one or two examples of each on the Octahedral lattice,
and our results on the Pyrochlore fit into the same categories.
One generic category is an ideal Luttinger-Tisza state, in which 
three degenerate ordering vectors are matched with three spin directions
to form a highly symmetric, already normalized state with no admixture of
other modes needed.  On the Octahedral lattice these were the 
two cuboctahedral states; our sole example on the Pyrochlore
is also cuboctahedral (Sec.~\ref{CS}).  The second category was
multiple-wavevector commensurate states, dominated by a star of symmetry-related
wavevectors but admixing others.  On the Octahedral lattice,
the only example was the ``double-twist''; on the Pyrochlore,
we have found {\it three} representatives:
the Kawamura states (Sec.~\ref{KawamuraStates}), 
a kind of Double-Twist state (Sec.~\ref{DoubleTwist}),
and a ``Multiply-Modulated Commensurate Spiral'' (M.M.C.S.) state
(Sec.~\ref{MMCS}).  The third category was
``conic spirals'' which were stackings of layers
with a uniform spin direction in each layer, 
which are built from two unrelated but nearly
degenerate spin modes, one of them being
generically incommensurate.
Being layered, these states can be optimized after
projecting all interactions onto a one-dimensional
``chain lattice''~\cite{SH}, as we outlined
in Sec.~\ref{sec:StacksAndColumns}.  
A corollary is that if a non-coplanar state 
is forced, the layers must be inequivalent, which is 
indeed the case for the (111) layering in which we did find
a conic spiral state (Appendix ~\ref{ACS}) in the Pyrochlore
lattice.  In both the Octahedral lattice and now in the
Pyrochlore lattice, it was difficult to stabilize
a conic spiral using the first few neighbor couplings:
in the Pyrochlore $J_3$ and $J_3'$ were needed.

\LATER{14 We need a paragraph commenting on how reliable this phase diagram
is.  In particular, how much have we been biased towards 
commensurate states by the necessity of doing iterative
minimization in rather small systems?  That can be illuminated,
I think, by comparing with the LT phase diagram.  I would have
guessed that, as you vary the couplings, the optimum Q point
often moves off the symmetry points and goes incommensurate.
But maybe that doesn't much happen with $J_1$-$J_2$ only; I'm
sure it happens when other couplings are added in.}

\LATER{14a A propos of the last remark, 
it might also be good to write a short paragraph commenting
how the LT phase diagram and actual phase diagram stack up.
There was a general hypothesis that we get states that mix
modes only when the modes are nearly degenerate -- is that
borne out, by the cases that had admixtures?}

When we examine the whole phase diagram, 
there seems to be a rough tendency that the wavevectors used for the
$(J_1,J_2)$ ground state are the same as those used for $(-J_1,-J_2)$;
for example, the Double-Twist state is diametrically opposite the 
Kawamura state in Figure~\ref{fig:PhaseDiag} and both
use optimal wavevectors of type 
$\left\{\tfrac{3}{4}\tfrac{3}{4}0\right\}$; the $\mb{q}=\mb{0}$ antiferromagnet is
opposite the ferromagnet which is (of course) also a $\mb{q}=\mb{0}$ state.
There is a plausible reason for this trend.  
Notice first that the matrix of $J_{ij}$'s (in real space)
has no diagonal terms, so its trace must be zero. 
The same must be true for its Fourier transform, so the sum of the four eigenvalues of $\un{\JLT}(\mb{q})$ is zero
at every wavevector.  The optimum wavevector $\QLT$ occurs
at the point in the zone where one branch of the eigenvalue spectrum
has its greatest positive excursion.  Necessarily,
the {\it sum} of the other three eigenvalues must have its
greatest {\it negative} excursion at the same wavevector: 
not uncommonly, that will be the point of the single
most negative excursion.  But if we reverse the signs of
all couplings, the L.T. matrix is the same except for a global
sign, and the greatest negative excursion becomes the
greatest positive one, i.e. the optimal mode.

\begin{acknowledgements}
We thank R. Z. Lamberty for discussions and advice regarding the numerical aspects of this project. This work was supported by NSF grant DMR-1005466.
\end{acknowledgements}

\appendix

\section{Fourier Transform of the Interactions for the Pyrochlore Lattice}
\label{App:LTmatrix}

In this appendix we present the matrix $\un{\JLT}(\mb{q})$ for the Pyrochlore lattice including interactions $J_1$, $J_2$, $J_3$ and $J_3'$ (i.e. including both kinds of third nearest neighbor interaction). One can calculate the elements of this matrix using equation \eqref{LTelements}. The matrix is:

	\beq
		\un{\mb{\JLT}}(\mb{q}) = J_1\un{\mb{M}_2} + J_2\un{\mb{M}_2} + J_3\un{\mb{M}_3} + J_3'\un{\mb{M}_3'}
	\eeq
where
\begin{widetext}
	\beq
		\un{\mb{M}_1} = 
		\begin{pmatrix}
			0 & \cos(\tfrac{q_y + q_z}{4}) & \cos(\tfrac{q_x + q_z}{4}) & \cos(\tfrac{q_x + q_y}{4}) \\
			\cos(\tfrac{q_y + q_z}{4})  & 0 & \cos(\tfrac{q_x - q_y}{4}) & \cos(\tfrac{q_x - q_z}{4}) \\
			\cos(\tfrac{q_x + q_z}{4})  & \cos(\tfrac{q_x - q_y}{4}) & 0 & \cos(\tfrac{q_y - q_z}{4})  \\
			\cos(\tfrac{q_x + q_y}{4})  & \cos(\tfrac{q_x - q_z}{4}) & \cos(\tfrac{q_y - q_z}{4})  & 0 
		\end{pmatrix} 
	\eeq

	\beq
		\un{\mb{M}_2} = 2
		\begin{pmatrix}
			0 & \cos(\tfrac{q_x}{2})\cos(\tfrac{q_y - q_z}{4}) & \cos(\tfrac{q_y}{2})\cos(\tfrac{q_z - q_x}{4}) & \cos(\tfrac{q_z}{2})\cos(\tfrac{q_x - q_y}{4}) \\
			\cos(\tfrac{q_x}{2})\cos(\tfrac{q_y - q_z}{4})  & 0 & \cos(\tfrac{q_z}{2})\cos(\tfrac{q_x + q_y}{4}) & \cos(\tfrac{q_y}{2})\cos(\tfrac{q_x + q_z}{4}) \\
			\cos(\tfrac{q_y}{2})\cos(\tfrac{q_z - q_x}{4})  & \cos(\tfrac{q_z}{2})\cos(\tfrac{q_x + q_y}{4}) & 0 & \cos(\tfrac{q_x}{2})\cos(\tfrac{q_y + q_z}{4})  \\
			\cos(\tfrac{q_z}{2})\cos(\tfrac{q_x - q_y}{4})  & \cos(\tfrac{q_y}{2})\cos(\tfrac{q_x + q_z}{4}) & \cos(\tfrac{q_x}{2})\cos(\tfrac{q_y + q_z}{4})  & 0 
		\end{pmatrix}
	\eeq
\end{widetext}

To concisely express the matrices that go with the two kinds of third nearest neighbor interaction, we introduce the condensed notation
	\begin{subequations}
	\beqa
		C_{xy} &=& \cos\left(\frac{q_x+q_y}{2}\right) \\
		\bar{C}_{xy} &=& \cos\left(\frac{q_x-q_y}{2}\right)
	\eeqa
	\end{subequations}
and a similar notation for cosine terms with arguments using $q_y$ and $q_z$.
\begin{widetext}
	\beq
		\un{\mb{M}_3} = 
		\begin{pmatrix}
		C_{xy} + C_{xz} + C_{yz} & 0 & 0 & 0 \\
		0 & \bar{C}_{xy} + \bar{C}_{xz} + C_{yz} & 0 & 0 \\
		0 & 0 & \bar{C}_{xy} + C_{xz} + \bar{C}_{yz} & 0 \\
		0 & 0 & 0 & C_{xy} + \bar{C}_{xz} + \bar{C}_{yz}
		\end{pmatrix}
	\eeq

	\beq
		\un{\mb{M}_3'} = 
		\begin{pmatrix}
		\bar{C}_{xy} + \bar{C}_{xz} + \bar{C}_{yz} & 0 & 0 & 0 \\
		0 & C_{xy} + C_{xz} + \bar{C}_{yz} & 0 & 0 \\
		0 & 0 & C_{xy} + \bar{C}_{xz} + C_{yz} & 0 \\
		0 & 0 & 0 & \bar{C}_{xy} + C_{xz} + C_{yz} 
		\end{pmatrix}
	\eeq
\end{widetext}

There are a few interesting things about this matrix. While $\un{\JLT}(\mb{q})$ is generally a hermitian matrix, it is symmetric in our case because every site on the Pyrochlore lattice is a center of inversion symmetry. Secondly, only $J_3$ and $J_3'$ terms appear on the diagonals since these are the only interactions that act between spins in the same F.C.C. sublattice. The $J_1$ and $J_2$ interactions only act between sites in different F.C.C. sublattices, so these show up only in off-diagonal terms. 

\section{A New Kind of Alternating Conic Spiral State}
\label{ACS}

Although no conic spiral-type ground states are found in the $J_1$-$J_2$ phase diagram of \eqref{HeisHam}, conic spiral ground states can be stabilized on the Pyrochlore lattice for certain sets of interactions. With the interactions $J_1 > 0$, $J_3 > 0$ and $J_3' < 0$, we find a new type of Alternating Conic Spiral state when the magnitude of $J_3'$ is much larger than the magnitudes of $J_1$ and $J_3$. Our state is distinct from the Alternating Conic Spiral found in Ref.~\onlinecite{SH} because the wavevector which controls the spiraling behavior is not parallel to the wavevector which controls the alternating behavior. This means that the component of the spins perpendicular to the conic axis spiral about that axis when you move in one direction in the lattice, and the component of the spins parallel to the conic axis alternates in sign when you move in a completely different direction in the lattice.

This state is made up of two different wavevectors, $\mb{q}_1$ and $\mb{q}_2$. The first wavevector $\mb{q}_1$ controls the spiraling behavior and it is likely that $\mb{q}_1$ does not have to be commensurate with the lattice (as is the case for the spiraling wavevector in the $\left\{q00\right\}$ planar spiral of section \ref{100spiral}). The second wavevector $\mb{q}_2$ controls the alternating behavior and in order for this state to be normalized it must be commensurate with the lattice and satisfy $\cos(\mb{q}_2\cdot\mb{r}) = \pm 1$ on all lattice sites $\mb{r}$. As mentioned above, $\mb{q}_1$ and $\mb{q}_2$ are not parallel.

We can express this state most concisely using one rotating basis vector. If we define the phase $\Phi = \mb{q}_1\cdot\mb{r}$, then we can write this basis vector as
	\beq
		\what{\mb{A}}(\mb{r}) = \cos(\Phi)\xhat + \sin(\Phi)\yhat \ .
	\eeq
With this rotating basis vector the parameterization of this state takes the form
	\begin{subequations}
	\label{AltCon}
	\beqa
		\mb{s}_0 &=& \sin\al \ \what{\mb{A}}(\mb{r}) + \cos\alpha\cos(\mb{q}_2\cdot\mb{r})\ \zhat \\
		\mb{s}_1 &=& \what{\mb{A}}(\mb{r}) \\
		\mb{s}_2 &=& \sin\al \ \what{\mb{A}}(\mb{r}) + \cos\alpha\cos(\mb{q}_2\cdot\mb{r})\ \zhat \\
		\mb{s}_3 &=& \cos(\mb{q}_2\cdot\mb{r})\ \zhat \ .
	\eeqa
	\end{subequations}

We can see from \eqref{AltCon} that the spins in F.C.C. sublattices 0 and 2 are performing Alternating Conic Spirals, the spins in F.C.C. sublattice 1 are performing a planar spiral about the conic axis, and the spins in F.C.C. sublattice 3 are alternating pointing parallel and antiparallel to the axis of the conic spiral.

An Iterative Minimization simulation at the parameter values $J_1 = 1$, $J_3 = 1$ and $J_3' = -4$ on a lattice of size $4\times 4\times 4$ with periodic boundary conditions found this new kind of Alternating Conic Spiral with $\mb{q}_1 = 2\pi(-\tfrac{3}{4},-\tfrac{1}{2},\tfrac{1}{4})$, $\mb{q}_2 = 2\pi(-\tfrac{1}{2},\tfrac{1}{2},\tfrac{1}{2})$ and cone angle $\al \approx 0.92$ (radians). As $J_3'$ is varied in this region, Fourier Transforms of the numerical spin configurations found in Iterative Minimization simulations still show large peaks near $\left\{\tfrac{1}{2}\tfrac{1}{2}\tfrac{1}{2}\right\}$ type wavevectors. For this reason, we suspect that the vector $\mb{q}_2$ is always a $\left\{\tfrac{1}{2}\tfrac{1}{2}\tfrac{1}{2}\right\}$ type wavevector (and therefore commensurate with the lattice). On the other hand, it appears that $\mb{q}_1$ and the cone angle $\al$ can vary continuously. 

\LATER{15 MFL: Should we say something about how future work could include 
variational optimization of this parameterization with respect to $\mb{q}_1$ and $\al$?}

For this set of parameter values, the optimal Luttinger-Tisza wavevector is of the type $\{\tfrac{1}{2}\tfrac{1}{2}\tfrac{1}{2}\}$, so $\mb{q}_2$ is equal to $\QLT$. The Luttinger-Tisza eigenvalue is $\LTeig = 7$. We find, however, that $\lam_{max}(\mb{q}_1) = 6.8629$, which is only slightly less than 7. So this is another example of a non-coplanar state which is composed of an optimal L.T. wavevector, $\mb{q}_2$, and a second type of wavevector, $\mb{q}_1$, which is only slightly sub-optimal. This provides further evidence for the hypothesis advanced in Ref.~\onlinecite{SH} that many non-coplanar states are made from combinations of optimal Luttinger-Tisza wavevectors and wavevectors that are only slightly suboptimal.


\section{Unit cells of the six sublattices of the Kawamura Sextuplet-q State}
\label{KawaSub}

These are the locations of lattice sites in the $2\times2\times2$ unit cells of the six sublattices for the example of the Kawamura Sextuplet-q state example presented above (section \ref{KawamuraSq}). This spin configuration was generated from an Iterative Minimization simulation with parameter values $J_1 = -1$, $J_2 = 0.1$ on a lattice of size $4\times4\times4$. 

\begin{enumerate}
\item F.C.C.0: (0,0,0), (1,1,1), (.5,0,1.5), (1.5,1,.5), (.5,1.5,0), (1.5,.5,1) \\ F.C.C.1: (.5,1.25,.75), (1.5,.25,1.75), (0,.75,.75), (1,1.75,1.75), (.5,.75,1.25), (1.5,1.75,.25) \\ F.C.C.2: (.25,1,.25), (1.25,0,1.25), (.75,1,1.75), (1.75,0,.75), (.25,.5,1.75), (1.25,1.5,.75)\\ F.C.C.3: (.25,.25,1), (1.25,1.25,0), (.25,1.75,.5), (1.25,.75,1.5), (.75,1.75,1), (1.75,.75,0)
\item F.C.C.0: (0,1,0), (1,1,0), (1.5,0,.5), (1.5,0,1.5), (.5,.5,1), (5,1.5,1) \\ F.C.C.1: (.5,.25,1.75), (.5,1.25,1.75), (0,1.75,.75), (1,1.75,.75), (1.5,.75,.25), (1.5,.75,1.25)\\ F.C.C.2: (.25,0,.25), (.25,0,1.25), (.75,1,.75), (1.75,1,.75), (1.25,.5,1.75), (1.25,1.5,1.75)\\ F.C.C.3: (1.25,.25,1), (1.25,1.25,1),(.25,.75,.5), (.25,.75,1.5), (.75,1.75,0), (1.75,1.75,0)
\item F.C.C.0: (0,.5,.5), (0,.5,1.5), (0,1.5,.5), (1,.5,1.5), (1,1.5,.5), (1,1.5,1.5)\\ F.C.C.1: (0,.25,.25), (0,.25,1.25), (0,1.25,.25), (1,.25,1.25), (1,1.25,.25), (1,1.25,1.25)\\ F.C.C.2: (.75,.5,1.25), (.75,1.5,.25), (.75,1.5,1.25), (1.75,.5,.25), (1.75,.5,1.25), (1.75,1.5,.25)\\ F.C.C.3: (.75,.25,1.5), (.75,1.25,.5), (.75,1.25,1.5), (1.75,.25,.5), (1.75,.25,1.5), (1.75,1.25,.5)
\item F.C.C.0: (0,1.5,1.5), (1,.5,.5)\\ F.C.C.1: (0,1.25,1.25), (1,.25,.25)\\ F.C.C.2: (.75,.5,.25), (1.75,1.5,1.25)\\ F.C.C.3: (.75,.25,.5), (1.75,1.25,1.5)
\item F.C.C.0: ((0,1,1), (1,0,0), (.5,0,.5), (1.5,1,1.5), (.5,.5,0), (1.5,1.5,1)\\ F.C.C.1: (.5,.25,.75), (1.5,1.25,1.75), (0,1.75,1.75), (1,.75,.75), (.5,.75,.25), (1.5,1.75,1.25)\\ F.C.C.2: (.25,1,1.25), (1.25,0,.25), (.75,0,.75), (1.75,1,1.75), (.25,1.5,1.75), (1.25,.5,.75)\\ F.C.C.3: (.25,1.25,1), (1.25,.25,0), (.25,1.75,1.5), (1.25,.75,.5), (.75,.75,0), (1.75,1.75,1)
\item F.C.C.0: (0,0,1), (1,0,1), (.5,1.5), (.5,1,1.5), (1.5,.5,0), (1.5,1.5,0)\\ F.C.C.1: (1.5,.25,.75), (1.5,1.25,.75), (0,.75,1.75), (1,.75,1.75), (.5,1.75,.25), (.5,1.75,1.25)\\ F.C.C.2: (1.25,1,.25), (1.25,1,1.25), (.75,0,1.75), (1.75,0,1.75), (.25,.5,.75), (.25,1.5,.75)\\ F.C.C.3: (.25,.25,0), (.25,1.25,0), (1.25,1.75,.5), (1.25,1.75,1.5), (.75,.75,1), (1.75,.75,1)
\end{enumerate}

\end{document}